\documentclass[12pt]{article} 
\usepackage[sectionbib]{natbib}
\usepackage{array,epsfig,fancyheadings,rotating}
\usepackage[]{hyperref}  
\usepackage{sectsty, secdot}
\sectionfont{\fontsize{12}{14pt plus.8pt minus .6pt}\selectfont}

\subsectionfont{\fontsize{12}{14pt plus.8pt minus .6pt}\selectfont}

\usepackage{amsmath}
\usepackage{amssymb}
\usepackage{amsfonts}
\usepackage{multirow}
\usepackage{amsthm}
\usepackage{authblk}
\usepackage{multirow}
\usepackage{lmodern}
\usepackage{bm}
\usepackage{pdflscape}
\usepackage{color}

\theoremstyle{definition}

\makeatletter
\@addtoreset{equation}{section}

\makeatother
\begin{document}


\renewcommand{\baselinestretch}{2}

\markright{ \hbox{\footnotesize\rm Statistica Sinica
}\hfill\\[-13pt]
\hbox{\footnotesize\rm
}\hfill }

\markboth{\hfill{\footnotesize\rm Shinichiro Shirota, Sudipto Banerjee and Alan E. Gelfand} \hfill}
{\hfill {\footnotesize\rm Spatial Joint Species Distribution Modeling} \hfill}

\renewcommand{\thefootnote}{}
$\ $\par


\fontsize{12}{14pt plus.8pt minus .6pt}\selectfont \vspace{0.8pc}
\centerline{\large\bf Spatial Joint Species Distribution Modeling}
\vspace{2pt} \centerline{\large\bf using Dirichlet Processes}
\vspace{.4cm} \centerline{Shinichiro Shirota$^{1}$, Alan E. Gelfand$^{2}$, Sudipto Banerjee$^{1}$} \vspace{.4cm} \centerline{\it
$^{1}$University of California, Los Angeles and $^{2}$Duke University} \vspace{.55cm} \fontsize{9}{11.5pt plus.8pt minus
.6pt}\selectfont

\begin{quotation}
\noindent {\it Abstract:}
Species distribution models usually attempt to explain presence-absence or abundance of a species at a site in terms of the environmental features (so-called abiotic features) present at the site.  Historically,  such models have considered species individually.  However, it is well-established that species interact to influence presence-absence and abundance (envisioned as biotic factors).  As a result, there has been substantial recent interest in joint species distribution models with various types of response, e.g., presence-absence, continuous and ordinal data.  Such models incorporate dependence between species response as a surrogate for interaction.

The challenge we address here is how to accommodate such modeling in the context of a large number of species (e.g., order $10^2$) across sites numbering on the order of $10^2$ or $10^3$ when, in practice, only a few species are found at any observed site.  Again, there is some recent literature to address this; we adopt a dimension reduction approach.  The novel wrinkle we add here is spatial dependence. That is, we have a collection of sites over a relatively small spatial region so it is anticipated that species distribution at a given site would be similar to that at a nearby site. Specifically, we handle dimension reduction through Dirichlet processes, enabling clustering of species, joined with spatial dependence across sites through Gaussian processes.

We use both simulated data and a plant communities dataset for the Cape Floristic Region (CFR) of South Africa to demonstrate our approach. The latter consists of presence-absence measurements for 639 tree species at 662 locations.  Through both data examples we are able to demonstrate improved predictive performance using the foregoing specification.

\vspace{9pt}
\par
\end{quotation}\par

\section{Introduction}\label{sec:I}
Understanding the distribution and abundance of species is a primary goal of ecological research.  In this regard, species distribution models are used to investigate the regressors that affect the presence-absence and abundance of species.  They can further be used to illuminate prevalence, predict biodiversity and richness, quantify species turnover, and assess response to climate change \citep{Midgleyetal(02), GuisanThuiller(05), Gelfandetal(06), Iversonetal(08), Botkinetal(07), McMahonetal(11), Thuilleretal(11)}.
These models are used to infer a species range either in geographic space or in climate space \citep{Midgleyetal(02)}, to identify and manage conservation areas \citep{AustinMeyers(96)}, and to provide evidence of competition among species \citep{Leathwick(02)}. A further key objective is interpolation, to predict species response at locations that have not been sampled.

Species distribution models (SDMs) are most commonly fitted to presence-absence data (binary) or abundance data (counts, ordinal classfications, or proportions). Occasionally, continuous responses are used such as biomass \citep{Dormannetal(12)}.
Prediction of species over space can be accommodated using a spatially explicit specification \citep{Gelfandetal(05a), Gelfandetal(06), Latimeretal(06)}.

Historically, SDMs have considered species individually \citep{Thuiller(03), Latimeretal(06), ElithLeathwick(09), Chakrabortyetal(11)}. To make predictions at the community scale, independent models for individual species are aggregated or stacked \citep{Calabreseetal(14)}. However, it is well-established that species interact to influence presence-absence and abundance. As a result, individual level models tend to predict too many species per location \citep{GuisanRahbek(11)}, as well as providing other misleading findings \citep[see][for some examples]{Clarketal(14)}.
Modeling species individually does not allow underlying joint relationships to be captured \citep{Clarketal(11), OvaskainenSoininen(11)}. Put differently, the problem can be viewed as the omission of the residual dependence between species.

Joint species distribution models (JSDMs) that incorporate species dependence include applications to presence-absence \citep{Pollocketal(14), Ovaskainenetal(10), OvaskainenSoininen(11)}, continuous or discrete abundance \citep{Latimeretal(09), Thorsonetal(15)}, abundance with large number of zeros \citep{Clarketal(14)} and recently, discrete, ordinal, and compositional data \citep{Clarketal(17)}. JSDMs jointly characterize the presence and/or abundance of multiple species at a set of locations, partitioning the drivers into two components, one associated with environmental suitability, the other accounting for species dependence through the residuals, i.e., adjusted for the environment.  Such models incorporate dependence between species response as a surrogate for attempting to supply formal specification of interaction.

JSDMs enhance understanding of the distribution of species, but their applicability has been limited due to computational challenges when there is a large number of species. To appreciate the potential challenge with presence-absence (binary) response and $S$ species, we have an $S$-way contingency table with $2^{S}$ cell probabilities at any given site. With observational data collection over space (and time), as in large ecological databases, the number of species is on the order of hundreds to thousands, rendering contingency table analysis infeasible. There is need for strategies to fit joint models in a computationally tractable manner.

To deal with these data challenges, we adopt dimension reduction techniques, working within  the Bayesian factor model setting \citep{West(03), LopesWest(04)}.
For instance, in the spatial case, \cite{RenBanerjee(13)} introduce spatial dependence into the factors using Gaussian predictive process models \citep{Banerjeeetal(08)}.
In our application, \cite{Taylor-Rodriguezetal(17)} also consider the dimension reduction within the factor modeling framework. They generate each row of the factor loading matrix from Dirichlet process realizations to enable common labels, i.e., clustering across the species. They assume independent factors because their plot locations are not close to each other.
Their focus is to jointly explain species presence at plots rather than predict the distribution at new locations. We add spatial dependence to the explanatory model to enable joint prediction at arbitrary locations over the study region.

In this regard, more recently, \cite{Thorsonetal(15)} implement spatial factor analysis for species distribution.  Their approach is to fix the factor loading matrix. \cite{Ovaskainenetal(16)} implement a multiplicative Gamma shrinkage prior proposed by \cite{BhattacharyaDunson(11)} for the factor loading matrix and introduce spatial dependence into the factors.  This work is the most comparable to our approach in the sense that both are specified through hierarchical models.  However, our specification directly models species dependence at the first (data) stage while  \cite{Ovaskainenetal(16)} bring dependence to the second (probabilities) stage.  We clarify this below.  Furthermore, our approach enables the data to inform about clustering among species.


We formulate such modeling in the context of a large number of species (e.g., order $10^2$) across a large number of sites (e.g., order $10^2$ or $10^3$) when, in practice, only a few species are found at any observed site.
Again, the novel wrinkle we add is spatial dependence. That is, we have a collection of sites over a relatively small spatial region so it is anticipated that species distribution at a given site would be similar to that at a nearby site.  As above, we adopt a dimension reduction approach, in particular, following modeling proposed by \cite{Taylor-Rodriguezetal(17)}. Specifically, we handle dimension reduction through Dirichlet processes, which enables joint labeling for species, i.e., clustering, joined with spatial dependence through Gaussian processes.

We use both simulated data and a plant communities dataset for the Cape Floristic Region (CFR) of South Africa to demonstrate our approach. The simulation study serves as a proof of concept for both continuous and binary response data.  The CFR dataset consists of presence-absence measurements for 639 tree species on 662 locations.  Through both data examples we are able to demonstrate improved predictive performance using the foregoing specification.

The format of the paper is as follows.
Section \ref{sec:SPM} introduces our motivating data and modeling strategy, i.e., spatial joint species distribution models with Dirichlet processes. Section \ref{sec:PS} provides the adaptation to binary responses along with discussion regarding identification of parameters specifically for probit models. In Section \ref{sec:BI}, we develop Bayesian inference for our model as well as our model comparison strategy. In Section \ref{sec:SS}, we investigate the proposed models with some simulation studies for continuous and binary response while in Section \ref{sec:RDA} we analyze the presence-absence data from the CFR. Finally, Section \ref{sec:SFW} offers discussion as well as potential future work.

\section{Spatial factor modeling with Dirichlet processes}\label{sec:SPM}
\subsection{A motivating data example}\label{sec:MDE}
Our data is extracted from a large database studying the distribution of plants in the Cape Floristic Region (CFR) of South Africa \citep{Takhtajan(86)}.  The CFR is one of the six floral kingdoms in the world and is located in the southwestern part of South Africa.  Though, geographically it is relatively small, it is extremely diverse ($9,000+$ species) and highly endemic ($70\%$ occur only in the CFR \citep{Rebelo(01)}.
There are more than $40,000$ sites with recorded sampling within the CFR.  The database from which our dataset was extracted consists of more than 1,400 plots with more than 2,800 species spanning six regions.
The data we use comes from one of these regions and exhibits high spatial clustering with $n=662$ plots and $S=639$ species. The response is binary, presence-absence for each species and plot (location).

\begin{figure}[ht]
 \begin{minipage}{0.46\hsize}
  \begin{center}
   \includegraphics[width=7cm]{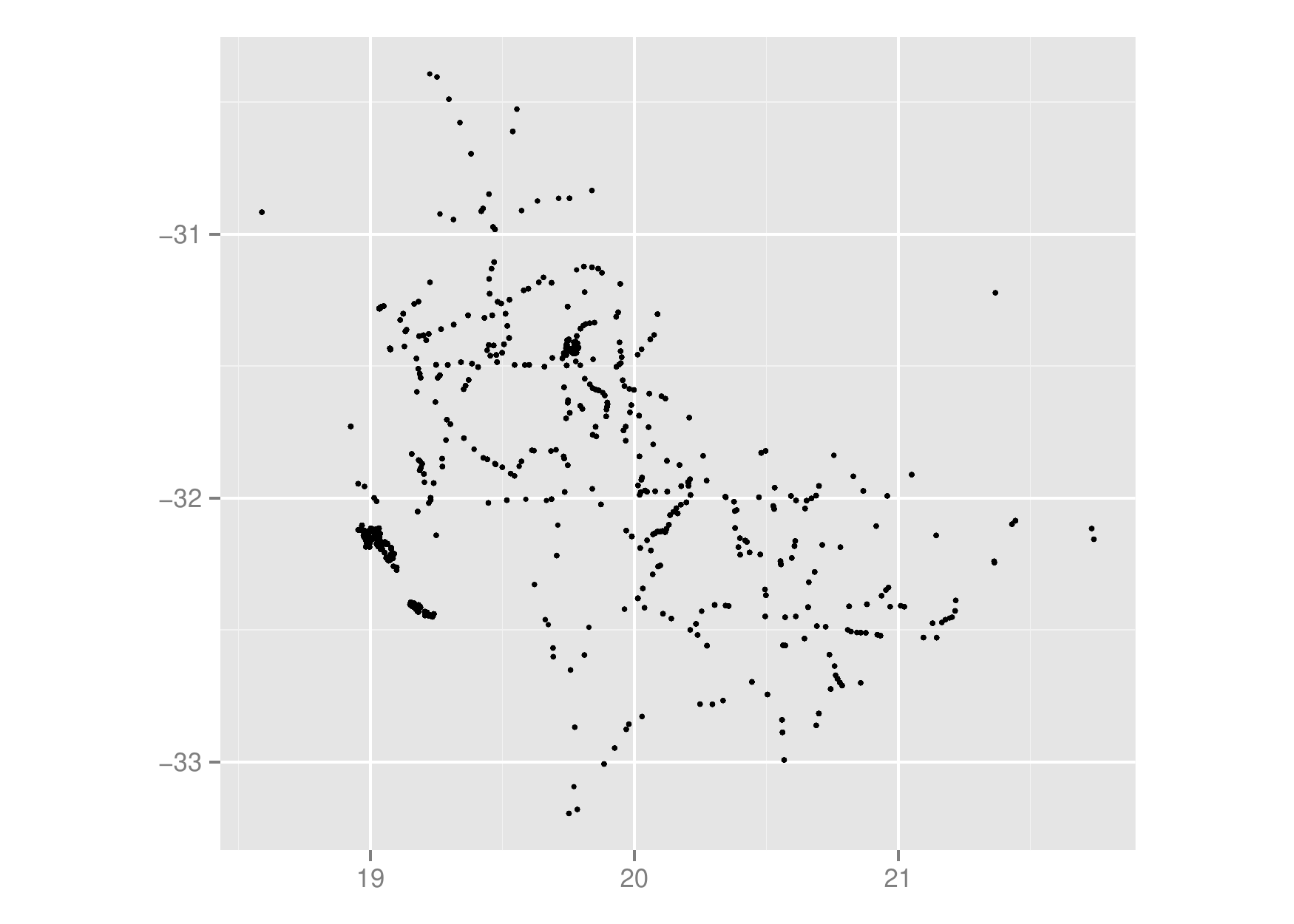}
  \end{center}
 \end{minipage}
 \hfill
 \hfill
 \begin{minipage}{0.50\hsize}
  \begin{center}
   \includegraphics[width=8cm]{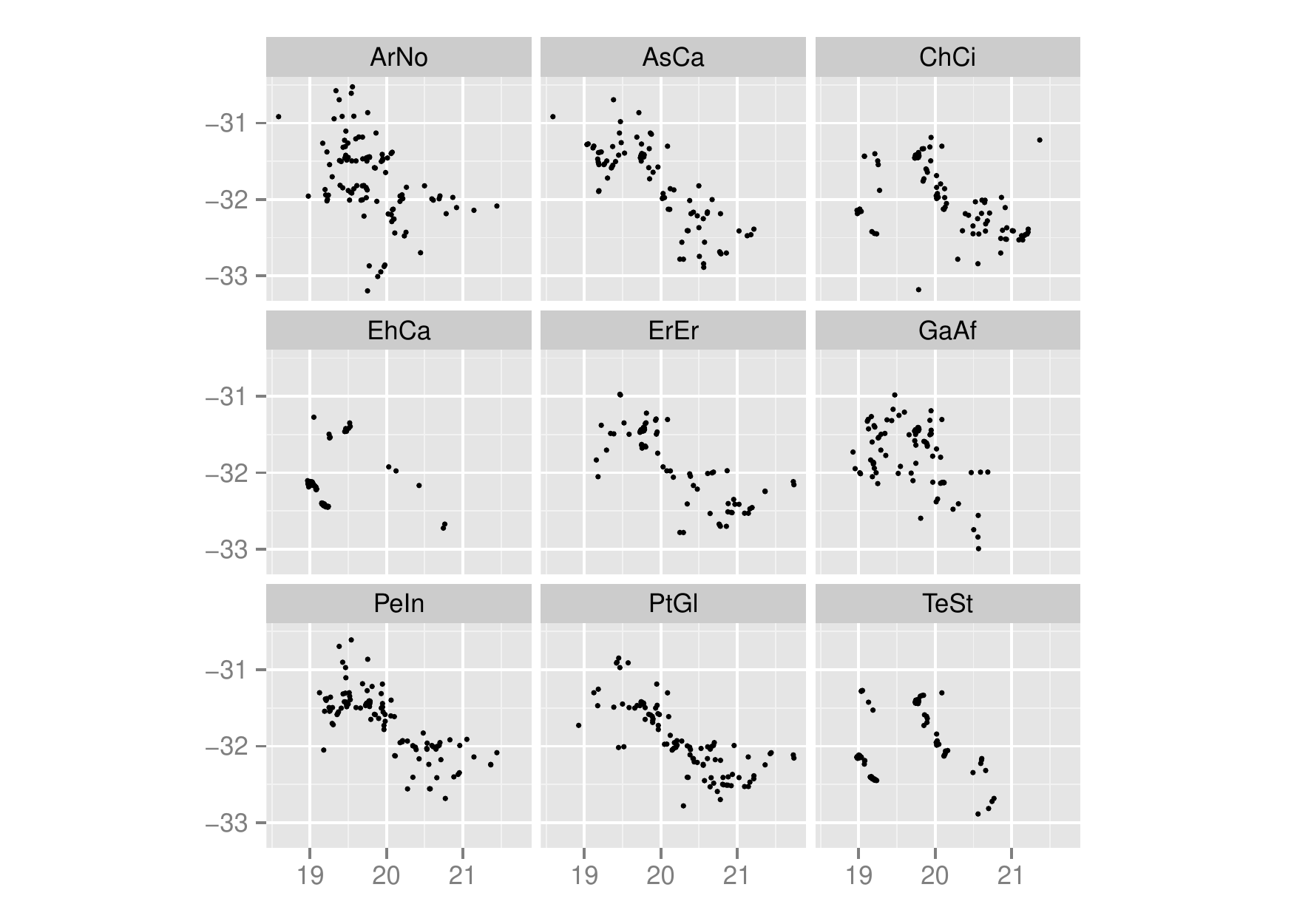}
  \end{center}
\end{minipage}
  \caption{662 locations in CFR (left) and the distribution of the presence of selected 9 species.}
  \label{fig:data}
\end{figure}

The left panel of Figure \ref{fig:data} shows the 662 locations in CFR data and the right panel shows the distribution of 9 selected species: 1) {\it Aridaria noctiflora} (ArNo); 2) {\it Asparagus capensis} (AsCa); 3) {\it Chrysocoma ciliata} (ChCi); 4) {\it Ehrharta calycina} (EhCa); 5) {\it Eriocephalus ericoides} (ErEr); 6) {\it Galenia africana} (GaAf); 7) {\it Pentzia incana} (PeIn); 8) {\it Pteronia glomerata} (PtGl); and 9) {\it Tenaxia stricta} (TeSt). These species are selected because they are observed on more than 100 locations (plots). Some species reveal strong spatial clustering, e.g., EhCa and TeSt.

Altogther, the total number of binary responses is $n\times S=662\times 639=423,018$. The overall number of presences is 6,980, 1.65$\%$ of the total number of binary responses.  This emphasizes the fact that, although we have many species in our dataset, only a few are present on any given plot. Among the $S=639$ species, 351 are observed in at most 5 locations. We discard these species and retain $S=288$ species across the $662$ locations for model fitting.

\subsection{Our model}\label{sec:OM}
Let $\mathcal{D}\subset \mathbb{R}^{2}$ be a bounded study region, $\mathcal{S}=\{\bm{s}_{1}, \ldots, \bm{s}_{n}\}$ be a set of plot locations where $\bm{s}_{i}\in \mathcal{D}$ for $i=1,\ldots,n$, and $\bm{U}_{i}:=\bm{U}(\bm{s}_{i})\in \mathbb{R}^{S}$ be an $S\times 1$ latent vector of continuous variables at location $\bm{s}_{i}$.
Under independence for the locations, the model for $\bm{U}_{i}$ is specified as
\begin{align}
\bm{U}_{i}=\mathbf{B}\bm{x}_{i}+\bm{\epsilon}_{i}, \quad \bm{\epsilon}_{i}\stackrel{iid}{\sim} \mathcal{N}_{S}(\bm{0}, \mathbf{\Sigma}), \quad \text{for} \quad i=1,\ldots,n \label{eq:(2.1)}
\end{align}
where $\mathbf{B}$ is an $S\times p$ coefficient matrix, $\bm{x}_{i}$ is a $p\times 1$ covariate vector at location $\bm{s}_{i}$ and $\mathbf{\Sigma}$ is a $S\times S$ covariance matrix for species.
This model has $\mathcal{O}(S^2)$ parameters, $S(S+1)/2$ parameters from $\mathbf{\Sigma}$ and $pS$ parameters from $\mathbf{B}$.
For example, for $S=300$ species and $p=3$ covariates, the model contains $46,050$ parameters.

\cite{Taylor-Rodriguezetal(17)} propose a dimension reduction approximation to $\mathbf{\Sigma}$ that allows the number of parameters to grow linearly in $S$.
They approximate $\mathbf{\Sigma}$ with $\mathbf{\Sigma}^{*}=\mathbf{\Lambda}\mathbf{\Lambda}^{T}+\sigma_{\epsilon}^2\mathbf{I}_{S}$ and replace the above model with
\begin{align}
\bm{U}_{i}=\mathbf{B}\bm{x}_{i}+\mathbf{\Lambda}\bm{w}_{i}+\bm{\epsilon}_{i}, \quad \bm{\epsilon}_{i}\sim \mathcal{N}_{S}(\bm{0}, \sigma_{\epsilon}^2 \mathbf{I}_{S}), \quad \text{for} \quad i=1,\ldots,n \label{eq:(2.2)}
\end{align}
where the random vectors $\bm{w}_{i}$ are i.i.d. with $\bm{w}_{i}\sim \mathcal{N}_{r}(\bm{0}, \mathbf{I}_{r})$ and $\mathbf{\Lambda}$ is an $S\times r$ matrix with $r\ll S$.
Now, $\mathbf{\Sigma}^{*}$ has only $Sr+1$ parameters, the estimation problem of $\mathcal{O}(S^2)$ parameters is reduced to that of $\mathcal{O}(S)$ parameters.  We refer to this specification as the dimension reduced nonspatial model.

Although $\mathbf{\Lambda}\mathbf{\Lambda}^{T}$ has rank $r$, including the nugget variance $\sigma_{\epsilon}^2\mathbf{I}$ ensures that $\mathbf{\Sigma}^{*}$ is nonsingular. The further approximation which \cite{Taylor-Rodriguezetal(17)} proposed is to sample the rows of $\mathbf{\Lambda}$ from a Dirichlet process mixture (DPM) using a stick-breaking representation \citep{Sethuraman(94)}.
The stick-breaking representation is attractive within a Gibbs sampling setting \cite[see, e.g.,][]{Escobar(94), EscobarWest(95), MacEachern(94), BushMacEachern(96), Neal(00)} due to a P\'{o}lya urn scheme representation which enables straightforward simulation from needed full conditional distributions.

Under the stick breaking construction, we say the random distribution, $G$, follows a DP with base measure $H$ and precision parameter $\alpha$, $G\sim DP(\alpha H)$, if $G(\cdot)=\sum_{l=1}^{\infty}p_{l}\delta_{\theta_{l}}(\cdot)$, where $p_{1}=\xi_{1}$, $p_{l}=\xi_{l}\prod_{h=1}^{l-1}(1-\xi_{h})$ $(h\ge 2)$ with i.i.d. $\xi_{l}\sim \text{Beta}(1,\alpha)$, and $\delta_{\theta_{l}}(\cdot)$ is the Dirac delta function at $\theta_{l}$ where $\theta_{l}\sim H$. Because it is almost surely a discrete distribution, this approach yields ties when realizations are drawn; the P\'{o}lya urn scheme representation draws from an atomic distribution having point masses at the already seen values with the remaining mass on $H$.  Thus, the DP enables us to model clustering.
We make use of this feature to allow some rows of $\mathbf{\Lambda}$ to be common, which corresponds to clustering species in their residual dependence behavior, as we clarify below.

According to (\ref{eq:(2.2)}), the $\bm{U}_{i}$ are conditionally independent given $\bm{B}$ and $\mathbf{\Lambda}$, i.e., the  $\bm{w}_{i}$ are independent across locations.
However, since the plot locations in our dataset are relatively close each other, we introduce spatial dependence into $\bm{w}_{i}$, which enables us to improve the prediction for new plot locations in the study region.

To provide the hierarchical formulation for this model, let $\mathbf{Z}=[\bm{Z}_{1} : \ldots : \bm{Z}_{N}]^{T}$ (with $\bm{Z}_{j}\sim H$) denote the $N\times r$ matrix whose rows make up all potential atoms. In this setup, we need a vector of grouping labels $\bm{k}=(k_{1},\ldots,k_{S})$ $(1\le k_{l}\le N)$ so that the $l$-th row of $\mathbf{\Lambda}$ is equal to $\bm{Z}_{k_{l}}$.  We note that $\mathbf{\Lambda}$ can be represented by $\mathbf{\Lambda}=\mathbf{Q}(\bm{k})\mathbf{Z}$ where $\mathbf{Q}(\bm{k})=[\bm{e}_{k_{1}}: \ldots : \bm{e}_{k_{S}}]^{T}$ is $S\times N$ with $\bm{e}_{k_{l}}$ denoting the $N$-dimensional vector with a 1 in position $k_{l}$ and $0$'s elsewhere. Letting $\mathbf{W}=[\bm{w}_{1}: \ldots : \bm{w}_{n}]^{T}$ be the $n\times r$ spatial factor matrix, our approximate model is
\begin{align}
\bm{U}_{i}|\bm{k}, \mathbf{Z}, \bm{w}_{i}, \mathbf{B}, \sigma_{\epsilon}^2 &\sim \mathcal{N}_{S}(\mathbf{B}\bm{x}_{i}+\mathbf{Q}(\bm{k})\mathbf{Z}\bm{w}_{i}, \sigma_{\epsilon}^2 \mathbf{I}_{S}), \quad \text{for} \quad i=1,\ldots, n, \nonumber \\
\mathbf{W}^{(h)}&\sim \mathcal{N}_{n}(\bm{0}, \mathbf{C}_{\phi}), \quad \text{for} \quad h=1,\ldots, r, \nonumber \\
k_{l}|\bm{p} &\sim \sum_{j=1}^{N}p_{j} \delta_{j}(k_{l}), \quad \text{for} \quad l=1,\ldots, S, \nonumber \\
\bm{Z}_{j}|\mathbf{D}_{\bm{Z}} &\sim \mathcal{N}_{r}(\bm{0}, \mathbf{D}_{\bm{Z}}), \quad \text{for} \quad j=1,\ldots, N, \label{eq:model} \\
Z_{1,h}&>0, \quad \text{for} \quad h=1,\ldots, r,  \nonumber \\
\bm{p}&\sim \mathcal{GD}_{N}(\bm{a}, \bm{b}), \nonumber \\
\mathbf{D}_{\bm{Z}}&\sim \mathcal{IW}(2+r-1, 4\text{diag}(1/\eta_{1}, \ldots, 1/\eta_{r})), \nonumber \\
\eta_{h}&\sim \mathcal{IG}(1/2, 1/10^4), \quad \text{for} \quad h=1,\ldots, r, \nonumber
\end{align}
where $\mathcal{GD}_{N}$ is an $N$ dimensional generalized Dirichlet distribution, $\mathbf{W}^{(h)}=(w_{1}^{(h)},\ldots, w_{n}^{(h)})^{T}$ is the $h$-th column of $\mathbf{W}$ ($n\times 1$ vector) and is distributed as an $n$-variate normal vector with mean $\bm{0}$ and covariance matrix $\mathbf{C}_{\phi}=[\exp(-\phi\| \bm{s}_{i}-\bm{s}_{i'} \|)]_{i,i'=1,\ldots, n}$, i.e., a realization of a Gaussian process (GP) with exponential covariance function at the sites in ${\cal S}$.  We refer to the above modeling specification as the dimension reduced spatial model. Again, \cite{Taylor-Rodriguezetal(17)} consider the entries in $\mathbf{W}^{(h)}$ to be independent across $i$ (i.e., across sites) while we introduce spatial dependence across $i$ through a GP for each column of $\mathbf{W}$. 
Furthermore, we restrict $k_{1}=1$ and all components of $\bm{Z}_{1}=(Z_{1,1},\ldots, Z_{1,r})^{T}$ to be positive in order to identify the covariance structure, as discussed in \cite{RenBanerjee(13)}. We provide more detail in Section \ref{sec:II}.

For prior specifications, we assume $\sigma_{\epsilon}^{2}\sim \mathcal{IG}(a/2, b/2)$ and $\mathbf{B}_{l}\sim \mathcal{N}(\bm{0}, c\mathbf{I}_{p})$ for $l=1,\ldots,S$ where $\mathbf{B}_{l}$ is $l$-th row of $\mathbf{B}$.
In practice, we suggest weakly informative prior specification, e.g., $a=2$ or $3$, $b\le 0.1$ and $c=100$.
We assume uniform prior for $\phi$, $\phi\sim \mathcal{U}[\phi_{min}, \phi_{max}]$ with $\phi_{max}=-\log(0.01)/d_{min}$ and $\phi_{min}=-\log(0.05)/d_{max}$ where $d_{max}$ and $d_{min}$ are the minimum and maximum distances across all the locations, following \cite{WangWall(03)}. 
In our datasets, $d_{max}=3.292$ and $d_{min}=0.0001$, so we obtain $\phi_{min}=0.909$ and $\phi_{max}=46,052$. 
Then, the induced {\it effective range} $d_{0}$, i.e., the distance at which spatial correlation is negligible (we set 0.05), is $d_{0}\in [0, 3.300]$ \citep{BanerjeeCarlinGelfand(14)}.

We offer a few clarifying remarks regarding the roles of $\Lambda$ and $\mathbf{w}_{h}$.

\noindent{\textbf{Remark 1:}} The initial specification in (\ref{eq:(2.2)}) is a nonspatial non-dimension reduced model.  The only model comparisons we make are between the dimension reduced nonspatial and spatial models since both of these models have the same approximation form for the covariance, $\mathbf{\Sigma}^{*}=\mathbf{\Lambda}\mathbf{\Lambda}^{T}+\sigma_{\epsilon}^2\mathbf{I}_{S}$.  In this regard, we would argue that $\mathbf{\Lambda}$ should not be location dependent. $\mathbf{\Lambda}\mathbf{\Lambda}^{T}$ is a feature of the taxonomy and should not be spatially varying.

\noindent{\textbf{Remark 2:}} We can clarify the interpretation of the clustering resulting from modeling the rows of $\mathbf{\Lambda}$ through a Dirichlet process.  If we are clustering on the rows of $\mathbf{\Lambda}$, then we are not clustering the species by their means since each species gets its own vector of regression coefficient from $\mathbf{B}$.  Rather, we are clustering on the residual covariance structure.  If row $\mathbf{\Lambda}_{l}=\mathbf{\Lambda}_{l'}$, then the row entries for $U_{i}^{(l)}$ and $U_{i}^{(l')}$ in $\mathbf{\Sigma}^*$ are identical.  In other words, when species are clustered at an iteration of the Markov chain Monte Carlo fitting, they have the same dependence structure with all other species.

So, the interpretation of posterior clustering for a pair of species is in terms of having similar dependence with all of the other species, adjusted for the regressors.  This may make useful ecological interpretation of the clustering difficult.  Alternatively, since attempting to formally model species interactions is very challenging, instead, we view modeling residual dependence as a surrogate. Then, we might attach an interpretation of similar dependence with other species as similar interaction with other species.

\noindent{\textbf{Remark 3:}} With regard to modeling the spatial dependence structure, in principle, each species might have its own spatial range/decay parameter.  However, under the dimension reduction we can include at most $r\ll S$ decay parameters.  So, an issue is whether incorporating a common decay parameter for the latent GP's, i.e.,  a separable model, will sacrifice much compared with employing $r$ decay parameters when $r$ is say 3 to 5.  The implications for the species level spatial dependence behavior are expected to be negligible.
Moreover, with $r$ decay parameters ordered \citep[as, e.g., in][]{RenBanerjee(13)} to obtain well-behaved Markov chain Monte Carlo (MCMC), the chain may not move well over this constrained space for the parameters.  Lastly, if we have an $S \times 1$ binary vector at each location, we would not expect the data to carry much information about a set of $r$ decay parameters.

\subsection{Interpretation}\label{sec:I}

Here we provide some technical elaboration of the foregoing remarks.  Given $\bm{w}_{i}$, the conditional expectations for $l$-th and $l'$-th row of $\bm{U}_{i}$ are
\begin{align}
E[U_{i}^{(l)}|\bm{w}_{i}]&=\mathbf{B}_{l}\bm{x}_{i}+\mathbf{\Lambda}_{l}\bm{w}_{i}\;\quad E[U_{i}^{(l')}|\bm{w}_{i}]=\mathbf{B}_{l'}\bm{x}_{i}+\mathbf{\Lambda}_{l'}\bm{w}_{i}
\end{align}
We see that the random effect provides an additional component in the mean explanation.  It is usually interpreted as capturing the effects of unmeasured/unobserved predictors at location $\bm{s}_{i}$. So, if we look at $\mathbf{\Lambda}_{l}\bm{w}_{i}$ and $\mathbf{\Lambda}_{l'}\bm{w}_{i}$, these inform about the residual variance adjusted for the fixed effects in the model. Also, we can study two features associated with the pair $\mathbf{\Lambda}_{l}\bm{w}_{i}$ and $\mathbf{\Lambda}_{l'}\bm{w}_{i}$. The first is the covariance between them which specifies the $(l,l')$-th entry in $\mathbf{\Lambda}\mathbf{\Lambda}^{T}$. The second is the expected distance between them, $E(\|\mathbf{\Lambda}_{l}\bm{w}_{i}-\mathbf{\Lambda}_{l'}\bm{w}_{i}\|^2)=(\mathbf{\Lambda}_{l}-\mathbf{\Lambda}_{l'})(\mathbf{\Lambda}_{l}-\mathbf{\Lambda}_{l'})^{T}$.

If $(\mathbf{\Lambda}_{l}-\mathbf{\Lambda}_{l'})(\mathbf{\Lambda}_{l}-\mathbf{\Lambda}_{l'})^{T}$ is small, this means we have multiple ties for the two species in their row selection in $\mathbf{\Lambda}$. So, for the two species, their residual random effects are similar, they provide similar residual adjustment. This is apart from whatever their mean contribution is. However, more importantly, it means that the pair have similar dependence structure with all of the remaining species. Evidently, when the $l$-th and $l'$-th row of $\mathbf{\Lambda}$ share the same cluster, $(\mathbf{\Lambda}_{l}-\mathbf{\Lambda}_{l'})(\mathbf{\Lambda}_{l}-\mathbf{\Lambda}_{l'})^{T}=\mathbf{O}$ (the matrix of zeros).  More generally, the labels do not change much across iterations in model fitting (see below) so $(\mathbf{\Lambda}_{l}-\mathbf{\Lambda}_{l'})(\mathbf{\Lambda}_{l}-\mathbf{\Lambda}_{l'})^{T}$ takes a discrete set of values for many pairs.

A different perspective makes the spatial random effects orthogonal to the fixed effects \citep[e.g.,][]{HodgesReich(10), HughesHaran(13), Hanksetal(15)}.  Let $\mathbf{X}=[\bm{x}_{1}:\ldots:\bm{x}_{n}]^{T}$ and $\mathbf{U}=[\bm{U}_{1}:\ldots:\bm{U}_{n}]^{T}$, $\mathbf{P}=\mathbf{X}(\mathbf{X}^{T}\mathbf{X})^{-1}\mathbf{X}^{T}$ be the projection matrix accociated $M(\mathbf{X})$, the column space spanned by $\mathbf{X}$. Then, we can write
\begin{align}
E[\mathbf{U}|\mathbf{W}]=\mathbf{X}\mathbf{B}^{T}+\mathbf{P}\mathbf{W}\mathbf{\Lambda}^{T}+(\mathbf{I}_{n}-\mathbf{P})\mathbf{W}\mathbf{\Lambda}^{T}
\end{align}
Thus, we can rewrite this conditional mean as
\begin{align}
E[\mathbf{U}|\mathbf{W}]&=\mathbf{X}\mathbf{B}^{*T}+\mathbf{W}^{*}\mathbf{\Lambda}^{T},
\end{align}
where $\quad \mathbf{B}^{*T} = \mathbf{B}^{T}+(\mathbf{X}^{T}\mathbf{X})^{-1}\mathbf{X}^{T}\mathbf{W}\mathbf{\Lambda}^{T}$ and $\mathbf{W}^{*}=(\mathbf{I}_{n}-\mathbf{P})\mathbf{W}$. This approach deals with {\it spatial confounding} which describes multicollinearity among spatial covariates $\mathbf{X}$ and spatial random effects $\mathbf{W}$. \cite{Paciorek(10)} demonstrated that this confounding can lead to bias in estimation, especially when the spatial random effects $\mathbf{W}$ are spatially smooth and have a large effective range of spatial autocorrelation. \cite{Hanksetal(15)} consider spatial confounding in the geostatistical (continuous spatial support) setting and demonstrate that the orthogonalization above provides computational benefits but its resulting Bayesian credible intervals can be inappropriately narrow under model misspecification.

In conclusion here, confounding is only a problem when interest lies in interpretation of the coefficient matrix, $\mathbf{B}$ rather than in prediction.  In particular, in our application below, Figures 7 and 8 reveal the difference in estimation between $\mathbf{B}$ and $\mathbf{B}^{*}$.  We anticipate that the ecological reader will care about the regressors and what role they play in the story when random effects are introduced, about how much confounding there is in the data and model.

\section{Adaptation to binary response, i.e., presence-absence data}\label{sec:PS}
For binary response data in the form of presence-absence, a logit or probit model specification is often assumed.
To work with binary responses, we adapt the data-augmentation algorithm proposed by \cite{ChibGreenberg(98)} for multivariate probit regression, which improves the mixing of the Markov chain Monte Carlo (MCMC) algorithm.
\cite{Taylor-Rodriguezetal(17)} consider the probit model specification,
\begin{align}
Y_{i}^{(l)}=\begin{cases}
             1 & U_{i}^{(l)}> 0 \\
             0 & U_{i}^{(l)}\le 0 \\
            \end{cases}, \quad \text{for} \quad l=1,\ldots, S, \quad i=1,\ldots,n \label{eq:(3.1)}
\end{align}
so that $U_{i}^{(l)}$ is an auxiliary variable.  We assume the modeling for $U_{i}^{(l)}$ as presented in Section \ref{sec:OM}. The form in (\ref{eq:(3.1)}) implies that we sample the latent $U_{i}^{(l)}$ from truncated normal distribution within MCMC iteration.

As a side remark, we specify that $Y_{i}^{(l)}=g(U_{i}^{(l)})=\mathbf{I}(U_{i}^{(l)}> 0)$.
The latent $U$s are part of the first stage model specification, i.e., $Y_{i}^{(l)}$ is a function of $U_{i}^{(l)}$.  The latent process driving the binary responses is specified at the data stage.
This contrasts with specifying a conditional distribution, $[Y_{i}^{(l)}|U_{i}^{(l)}]$, e.g., $P(Y_{i}^{(l)}=1) = p(U_{i}^{(l)})$ where $p(\cdot)$ would be a regression in $U_{i}^{(l)}$, e.g., $\Phi(\alpha_0 + \alpha_{1} U_{i}^{(l)})$.
This moves the $U$'s to a second stage model specification and would also yield a probit regression.

To add some clarification, the former says that the $Y_{i}^{(l)}$ arises deterministically from the $U_{i}^{(l)}$ surface.  The latter says we have a Bernoulli trial with a probit link function at each $i$.  It is not clear that the former is better than the latter.  Perhaps it might be preferred because you are directly modeling the dependence, joint and spatial, between $U_{i}^{(l)}$ and $U_{i'}^{(l')}$, hence between $Y_{i}^{(l)}$ and $Y_{i'}^{(l')}$, rather than deferring the dependence to the second stage, i.e., to the presence absence surface with conditionally independent Bernoulli trials at each location given the surface.
Again, this is the distinction between our approach and that of Ovaskainen et al. (2016).

\subsection{Identifiability issues}\label{sec:II}
We seek to learn about the  dependence structure between species through $\mathbf{\Sigma}^{*}=\mathbf{\Lambda}\mathbf{\Lambda}^{T}+\sigma_{\epsilon}^{2}\mathbf{I}_{S}$ as well as to extract clustering behavior for the rows of $\mathbf{\Lambda}$.  However, it is well known that, with random $\mathbf{W}$, the entries in $\mathbf{\Lambda}$ and $\sigma_{\epsilon}^{2}$ are not identified.  So, we briefly review the identification problems involved in factor models and probit models. The identifiability problems for each of these specifications are mutually connected.

First, consider the factor loading matrices and factor vectors under the dimension reduction. For posterior inference, we identify $\mathbf{\Lambda}\bm{w}$ but not $\mathbf{\Lambda}$ and $\bm{w}$. Some restriction on the factor loading matrices is required \citep{GewekeSingleton(80), LopesWest(04)}.
A widely used approach is to fix certain elements of $\mathbf{\Lambda}$, usually to zero, such as restricting $\mathbf{\Lambda}$ to be upper or lower triangular matrices with strictly positive diagonal elements \citep{GewekeZhou(96)}.
This restriction enables direct interpretation of latent factors and loading matrices.

Alternatively, \cite{RenBanerjee(13)} discuss the difference with regard to identifiability according to whether the elements in factor vectors across locations ($\mathbf{W}^{(h)}$ for $h=1,\ldots, r$) are independent or are spatially structured across locations.  In the former case, dependence structure is invariant to any orthogonal transformation of $\mathbf{\Lambda}$.  We can have an infinite number of equivalent matrices of factor loadings.  However, in the second case, they argue that only two types of linear transformations, reflections and permutations, lead to non-identifiability. In order to avoid these types of non-identifiability, \cite{RenBanerjee(13)} put a positive restriction on the first row of $\mathbf{\Lambda}$.
This is available for our modeling as well, but does not impose constant constraints on $\mathbf{\Lambda}$ so the elements of $\mathbf{\Lambda}$ and $\bm{w}$ themselves still cannot be identified.
However, the restrictions suggested by \cite{RenBanerjee(13)} enable us to identify the covariance structure of the latent process, i.e., $\text{Cov}[\text{vec}(\mathbf{U})]$, which is one of our goals.

\section{Bayesian inference}\label{sec:BI}
\subsection{Model fitting}
The full joint likelihood is
\begin{align}
\mathcal{L}&\propto (\sigma_{\epsilon}^2)^{(nS/2+1)}\prod_{i=1}^{n}\exp\biggl(-\frac{1}{2\sigma_{\epsilon}^2}\|\bm{U}_{i}-\mathbf{B}\bm{x}_{i}-\mathbf{Q}(\bm{k})\mathbf{Z}\bm{w}_{i}\|^2 \biggl) \nonumber \\
&\times |\mathbf{C}_{\phi}|^{-1/2} \prod_{h=1}^{r} \exp\biggl(-\frac{1}{2}\mathbf{W}^{(h)T}\mathbf{C}_{\phi}^{-1} \mathbf{W}^{(h)}\biggl)\mathcal{IG}\biggl(\sigma_{\epsilon}^{2}|\frac{a}{2}, \frac{b}{2}\biggl)\prod_{l=1}^{S}\mathcal{N}(\mathbf{B}_{l}|\bm{0}, c\mathbf{I}_{p}) \nonumber \\
&\times |\mathbf{D}_{\bm{Z}}|^{-1/2}\prod_{j=1}^{N}\exp\biggl(-\frac{1}{2}\bm{Z}_{j}^{T}\mathbf{D}_{\bm{Z}}^{-1}\bm{Z}_{j}\biggl)\times \prod_{l=1}^{S}\sum_{j=1}^{N}p_{j}\delta_{j}(k_{l})\pi(\bm{p}|\bm{0}, \bm{\alpha}) \nonumber \\
&\times \mathcal{IW}\biggl(\mathbf{D}_{\bm{Z}}|2+r-1, 4\text{diag}\biggl(\frac{1}{\eta_{1}}, \ldots, \frac{1}{\eta_{r}}\biggl)\biggl)\prod_{h=1}^{r}\mathcal{IG}\biggl(\eta_{h}|\frac{1}{2}, \frac{1}{10^4}\biggl)\mathcal{U}(\phi|\phi_{min}, \phi_{max})
\end{align}
Our sampling algorithm is similar to that of \cite{Taylor-Rodriguezetal(17)} except for sampling $\mathbf{W}$ and $\phi$. In our case, $\mathbf{W}$ has spatial correlation, but Gibbs sampling is still available.  We describe the full sampling steps including sampling of  $\mathbf{W}$ and $\phi$ in the Appendix.
\\

\subsection{Model comparison}\label{sec:MC}

Our focus for model comparison is with regard to improvement of the predictive performance at held out locations.
We implement out-of-sample predictive performance checks with respect to held out samples of entire plots rather than holding out samples of species within plots. This is in accord with our spatial modelling objective, to improve  predictive performance for held out locations.

For the continuous response case, predictive performance is assessed by calculating the Euclidean distances between the true values and the conditional predictions, predicting $100 p\%$ of the plots, conditional on the remaining $100 (1-p)\%$ plots.
We denote the number of plots of test data by $m$ and the out-of-sample response matrix (test data) by $\mathbf{U}_{pred}=(\bm{U}_{1,pred},\ldots, \bm{U}_{m,pred})$ at locations $\mathcal{S}_{pred}=\{\bm{s}_{i_{1}},\ldots, \bm{s}_{i_{m}} \}$.

The criterion used to assess predictive ability of the algorithm is the predictive mean squared error (PMSE), given by
\begin{align}
\text{PMSE}=\frac{1}{Sn_{p}} \sum_{i=1}^{m}(\bm{U}_{i, pred}-\hat{\bm{U}}_{i, pred})^{T}(\bm{U}_{i, pred}-\hat{\bm{U}}_{i, pred})
\end{align}
where $\hat{\bm{U}}_{i, pred}$ is the posterior mean estimate of $\bm{U}_{i, pred}$.

For binary responses, we use the Tjur $R^2$ coefficient of determination \citep{Tjur(09)}, which compares the estimated probabilities of presence between the observed ones and the observed zeros. For species $j$, this quantity is given by $TR_{j}=(\hat{\pi}_{j}(1)-\hat{\pi}_{j}(0))$ where $\hat{\pi}_{j}(1)$ and $\hat{\pi}_{j}(0)$ are the average probabilities of presence for the observed ones and zeros of the $j$-th species across the locations, respectively.
The larger the $TR_{j}$ , the better the discrimination.
We calculate an average TR measure across species, i.e., $TR=\frac{1}{S}\sum_{j=1}^{S}TR_{j}$.
\section{A simulation study}\label{sec:SS}

\subsection{Continuous responses}
We investigate the parameter recovery of our proposed model for continuous responses.
We use the same locations $(n=662)$ and covariate information as in the CFR data.
As covariate information, we include: (1) elevation, (2) mean annual precipitation, and (3) mean annual temperature; these values are standardized.
The setting for the simulated data is
\begin{align}
q&=5, \quad p=3, \quad S=300, \quad K_{true}=10, \quad \sigma_{\epsilon}^2=1 \nonumber \\
\bm{U}_{i}&\sim \mathcal{N}(\tilde{\mathbf{B}}\bm{x}_{i}+\mathbf{Q}_{true}(\bm{k})\mathbf{Z}_{true}\bm{w}_{i}, \sigma_{\epsilon}^2 \mathbf{I}_{S}), \quad i=1,\ldots, n  \nonumber \\
\tilde{\mathbf{B}}_{l}&\sim \mathcal{N}(\bm{0}, \mathbf{I}_{p}), \quad l=1,\ldots, S \\
\mathbf{W}^{(h)}&\sim \mathcal{N}(\bm{0}, \mathbf{C}_{\phi}), \quad h=1,\ldots, q \nonumber \\
\mathbf{Z}_{true}&=(\bm{Z}_{1,true},\ldots,\bm{Z}_{K_{true},true})^{T}. \nonumber
\end{align}
Here, $q$ denotes the fixed number of factors under the simulation.  $\mathbf{W}^{(h)}$ is $h$-th column of $\mathbf{W}$, an $n$-variate normal vector with mean $\bm{0}$ and covariance matrix $\mathbf{C}_{\phi}=[\exp(-\phi \| \bm{s}_{i}-\bm{s}_{i'} \|)]_{i,i'=1,\ldots, n}$, we set $\phi=2$.
The label $k_{l}$ is uniformly sampled from $K_{true}$ labels for $l=1,\ldots, S$.
$\mathbf{Q}_{true}(\bm{k})$ and $\mathbf{Z}_{true}$ are $S\times K_{true}$ and $K_{true}\times q$ matrices, respectively.
Each component of $\bm{Z}_{k,true}$ is uniformly selected from $\{-1,-0.5,0,0.5,1\}$, e.g., a realization might be $\bm{Z}_{k,true}=(0.5, -0.5,0,0,1)^{T}$, so that $\bm{Z}_{k,true}\neq \bm{Z}_{k',true}$ for $k<k'=1,\ldots,K_{true}$ and we set $\bm{Z}_{1,true}=0.5\bm{1}_{q}$.
We forced the $\bm{Z}_{k,true}$ to be quite different from each other in order to facilitate recovery of the number of clusters, especially for the binary case. We set $\bm{Z}_{1,true}=0.5\bm{1}_{q}$ to keep all components of $\bm{Z}_{1,true}$ positive in order to meet the identifiability condition discussed in Section \ref{sec:II}.

We estimate posterior for $\tilde{\mathbf{B}}, \mathbf{Z}, \mathbf{W}, \bm{k}, \sigma_{\epsilon}^2, \phi$ through Bayesian inference, with model fitting described in appendix \ref{app:A1}.
The prior specification is
\begin{align}
\sigma_{\epsilon}^{2}\sim \mathcal{IG}(2, 0.1), \quad \phi\sim \mathcal{U}[\phi_{min}, \phi_{max}], \quad \tilde{\mathbf{B}}_{l}&\sim \mathcal{N}(\bm{0}, 100\mathbf{I}_{p}), \quad \text{for} \quad l=1,\ldots, S
\end{align}
where $\phi_{min}=0.909$ and $\phi_{max}=46,052$.
We adopt dimension reduction selecting $r=5$ and $N=150$ ($>K_{true}$ and $<S$).
We run the MCMC, discarding the first 20,000 samples as a burn-in period, preserving the subsequent 20,000 samples as posterior samples.

Table \ref{tab:SimC-result} provides the estimation results for our model fitting.
Both the decay parameter $\phi$ and the nugget variance $\sigma_{\epsilon}^2$ are well recovered.

\begin{table}[ht]
\caption{Estimation results for continuous response}
\label{tab:SimC-result}
\centering
\begin{tabular}{lcccccc}
\hline
\hline
 & True & Mean & Stdev & $95\%$ Int  \\
\hline
$\phi$ & 2 & 2.095 & 0.226 & [1.600, 2.585]  \\
$\sigma_{\epsilon}^2$ & 1 & 1.000 & 0.003 & [0.993, 1.006]   \\
\hline
\hline
\end{tabular}
\end{table}

\begin{figure}[ht]
  \begin{center}
   \includegraphics[width=12cm]{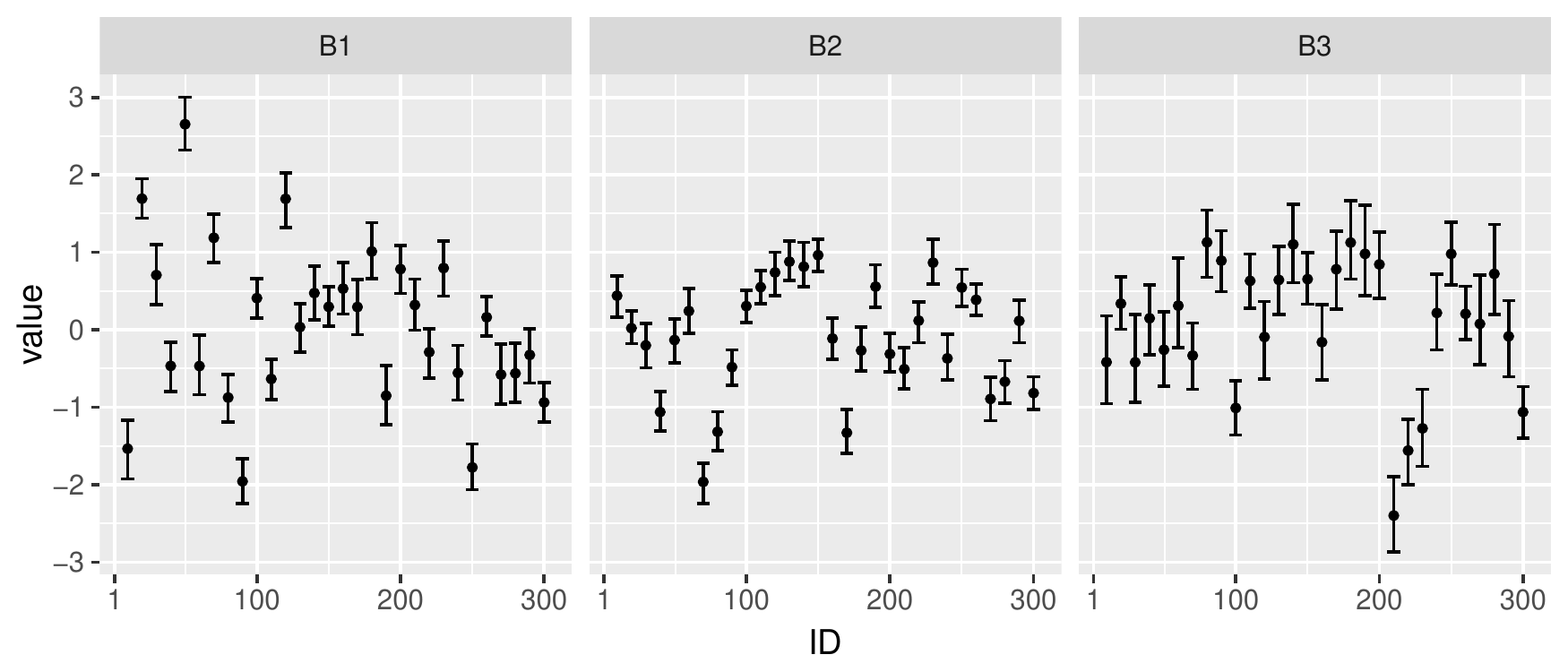}
  \end{center}
  \caption{Estimated 95$\%$ CIs of $\tilde{\mathbf{B}}$ with continuous responses for 30 selected species. Black dots denote the true values. }
  \label{fig:SimC-B}
\end{figure}

\begin{figure}[ht]
  \begin{center}
   \includegraphics[width=8cm]{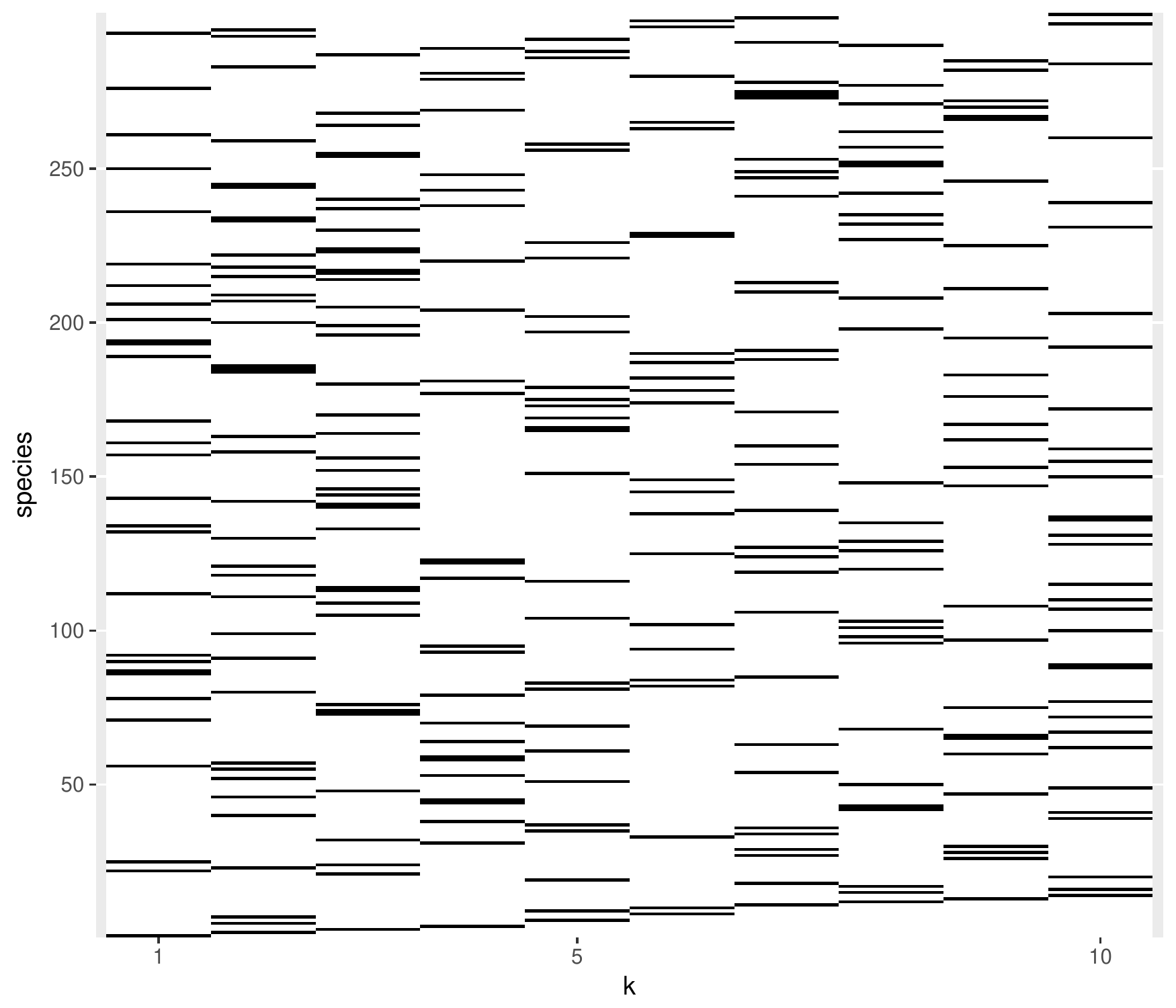}
  \end{center}
  \caption{For continuous response, the 0-1 map (0:white, 1:black) of sampled $\bm{k}$ for the spatial model with maximum posterior probability. Each species has only one label.}
  \label{fig:SimC-K}
\end{figure}

Figure \ref{fig:SimC-B} shows the $95\%$ credible intervals (CIs) for $\tilde{\mathbf{B}}$ for 30 selected species (chosen every 10 species) by our model.
With $\tilde{\mathbf{B}}$ identified in the case of continuous response, the true parameter values are well recovered for both cases.
Figure \ref{fig:SimC-K} reveals the sampled $\bm{k}$ of our spatial model for all species with maximum posterior probability.
Indeed, in this simulation study, $\bm{k}$s for both models are completely recovered.
In other words, the number of components of $\bm{k}$ is 10 ($=K_{true}$) with posterior probability 1 for both independence and spatial models.
The sampled $\bm{k}$s for both models are also the same as simulated $\bm{k}$ with posterior probability 1.

In addition, we compare the true covariance $\mathbf{\Sigma}^{*}=\mathbf{\Lambda}\mathbf{\Lambda}^{T}+\sigma_{\epsilon}^{2}\mathbf{I}_{S}$ with the estimated covariance $\hat{\mathbf{\Sigma}}^{*}=\hat{\mathbf{\Lambda}}\hat{\mathbf{\Lambda}}^{T}+\hat{\sigma}_{\epsilon}^{2}\mathbf{I}_{S}$ where $\hat{\mathbf{\Lambda}}$ and $\hat{\sigma}_{\epsilon}^{2}$ are posterior means of $\mathbf{\Lambda}$ and $\sigma_{\epsilon}^{2}$ under the spatial and independent models.
This comparison is motivated by the possibility that, with dependence in the spatial factors, the estimated covariance structure might be distorted assuming independent factors.
We calculate the Frobenius norm, i.e., $\|\mathbf{A}\|_{F}=\sqrt{\sum_{l=1}^{S}\sum_{l'=1}^{S}|a_{ll'}|^2 }$, for the difference $\mathbf{\Sigma}^{*}-\hat{\mathbf{\Sigma}}^{*}$.
The values are 161.8 for the independent model and 31.13 for the spatial model.
Hence, when factors have spatial dependence, the independence model appears to provide less precise estimation of  $\hat{\mathbf{\Sigma}}^{*}$.

Finally, we investigate the predictive performance of our spatial model.
As discussed in Section \ref{sec:MC}, the predictive performance is assessed by calculating the Euclidean distances between the true values and the conditional predictions, predicting $20\%$ of the plots, conditional on observing the remaining $80\%$ of the plots.
The estimated PMSE for our spatial model is 1.144 and that for the independence model is 2.069.
The spatial model reveals a roughly $45\%$ improvement over the independence model.


\subsection{Binary responses}
In addition to the continuous case, we also investigate the parameter recovery and the estimated covariance structure for binary responses.
In the binary case, all parameter settings are the same as in the continuous case except for the observed response,
\begin{align}
Y_{i}^{(l)}&=\begin{cases}
         1, \quad U_{i}^{(l)}>0 \\
         0, \quad U_{i}^{(l)}\le 0
\end{cases}, \quad i=1,\ldots, n, \quad l=1,\ldots, S.
\end{align}
We sample $\mathbf{U}$ as auxiliary responses within MCMC iterations.
Again, we discard the first 20,000 samples as burn-in period and preserve the subsequent 20,000 samples as posterior samples. The same prior specification is assumed for $\phi$ and $\tilde{\mathbf{B}}$ and we fix $\sigma_{\epsilon}^{2}=1$.
The posterior mean of $\phi$ is 1.687 ($95\%$ CI [1.237, 2.422]) so the true value is well recovered.

\begin{figure}[ht]
  \begin{center}
   \includegraphics[width=12cm]{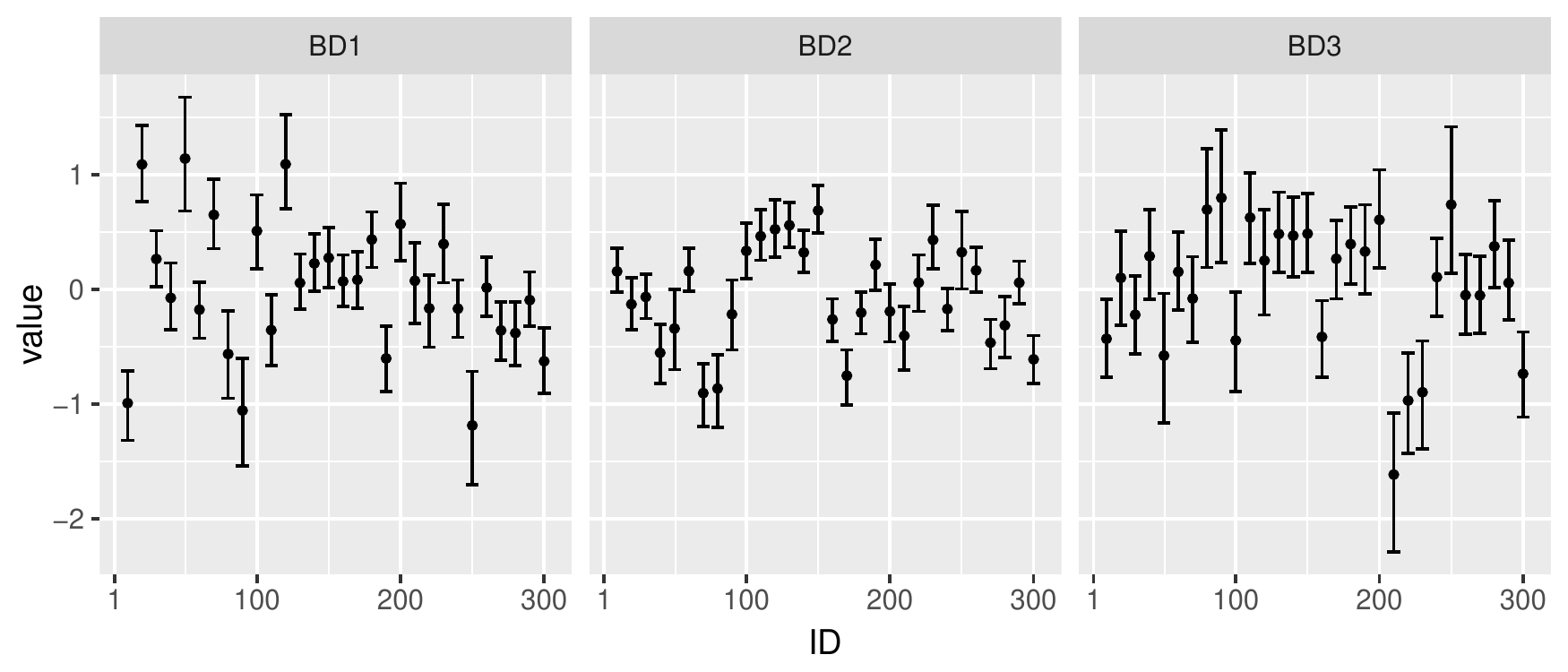}
  \end{center}
  \caption{Estimated 95$\%$ CIs of $\mathbf{D}_{\mathbf{\Sigma}}^{-1/2}\tilde{\mathbf{B}}$ with binary response for 30 selected species. Black dots denote the true values. }
  \label{fig:SimB-B}
\end{figure}

For the binary case, $\tilde{\mathbf{B}}$ is not identifiable.  \cite{Taylor-Rodriguezetal(17)} estimate $\mathbf{B}$ with a scaled correlation matrix, $\mathbf{R}=\mathbf{D}_{\mathbf{\Sigma}^{*}}^{-1/2}\mathbf{\Sigma}^{*}\mathbf{D}_{\mathbf{\Sigma}^{*}}^{-1/2}$, i.e., $\mathbf{B}=\mathbf{D}_{\mathbf{\Sigma}^{*}}^{-1/2}\tilde{\mathbf{B}}$, following the discussion in \cite{Lawrenceetal(08)}.
We adopt this choice as well because applying the change of variables $(\tilde{\mathbf{B}}, \mathbf{\Sigma}^{*})$ to $(\mathbf{B}, \mathbf{R})$ does not affect the probabilities for $\bm{Y}_{i}$ but identifies $\mathbf{B}$ to be unaffected by the change of scale matrix, $\mathbf{D}_{\mathbf{\Sigma}^{*}}$.
Figure \ref{fig:SimB-B} shows the $95\%$ CIs for $\mathbf{D}_{\mathbf{\Sigma}}^{-1/2}\tilde{\mathbf{B}}$ for 30 selected species (chosen every 10 species) under our model.
The true parameter values are well recovered.

\begin{figure}[ht]
  \begin{center}
   \includegraphics[width=8cm]{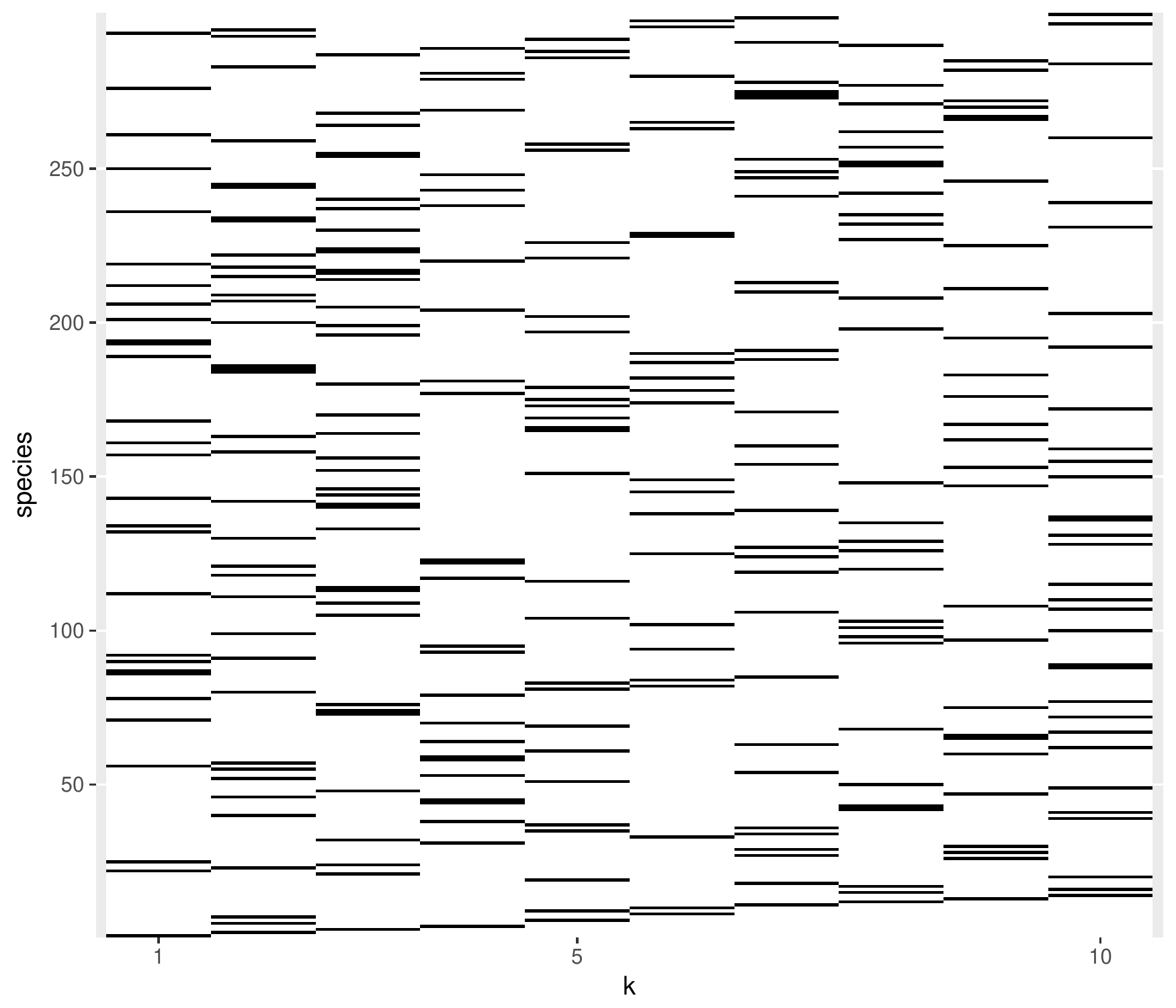}
  \end{center}
  \caption{For binary response, the 0-1 map (0:white, 1:black) of sampled $\bm{k}$ for the spatial model with maximum posterior probability. Each species has only one label.}
  \label{fig:SimB-K}
\end{figure}

Figure \ref{fig:SimB-K} shows the 0-1 map of the sampled $\bm{k}$ for the spatial model with maximum posterior probability.
As in the continuous case, $\bm{k}$ is completely recovered, i.e., the estimated number of clusters is 10 with posterior probability 1, and $\bm{k}$ is the same as true $\bm{k}$ with posterior probability 1 after a sufficiently long burn-in period.

Again, we compare the true covariance $\mathbf{\Sigma}^{*}=\mathbf{\Lambda}\mathbf{\Lambda}^{T}+\mathbf{I}_{S}$ and the estimated covariance $\hat{\mathbf{\Sigma}}^{*}=\hat{\mathbf{\Lambda}}\hat{\mathbf{\Lambda}}^{T}+\mathbf{I}_{S}$ for the spatial and independent models.
The calculated Frobenius norms are 156.1 for the independent model and 73.09 for the spatial model.
The value for the spatial model is smaller than that of the independent model but larger than that of the spatial model with continuous responses. Finally, we investigate the predictive performance of our spatial model using the TR measures introduced in Section \ref{sec:MC}.  The values are 0.5603 for the spatial model and 0.415 for the independent model; the spatial model outperforms the independent model.

\section{Real data application}\label{sec:RDA}
From Section \ref{sec:MDE}, the total number of binary responses is $n\times S=662\times 639=423,018$.
The number of $Y_{l,i}=1$ is 6,980, 1.65$\%$ of all binary responses.
Discarding the 351 species that are observed at at most 5 locations, we preserve $S=288$ species for model fitting.
Longitude and latitude are transformed into easting and northing scales.
Then, these scales are normalized by 100 km, so $\| \bm{s}_{i}-\bm{s}_{i'} \|=1$ means the distance between $\bm{s}_{i}$ and $\bm{s}_{i'}$ is 100 km.
Again, as covariate information, we include: (1) elevation, (2) mean annual precipitation, (3) mean annual temperature; again, these values are standardized.

In the analysis below, we set $r=5$ \citep[following][]{Taylor-Rodriguezetal(17)}.  (We conducted some sensitivity analysis with regard to the choice of $r$, see below.)
The prior specification is
\begin{align}
\phi\sim \mathcal{U}[\phi_{min}, \phi_{max}], \quad \mathbf{B}_{l}&\sim \mathcal{N}(\bm{0}, 100\mathbf{I}_{p}), \quad \text{for} \quad l=1,\ldots, S
\end{align}
where $\phi_{\min}=0.909$ and $\phi_{max}=46,052$ and we fix $\sigma_{\epsilon}^{2}=1$
We discard the first 20,000 samples as burn-in period and preserve the subsequent 20,000 samples as posterior samples.

The estimated value of $\phi$ is 2.314 ($95\%$ CI [1.614, 3.589]), which reflects the spatial dependence for the factors.
Among 288 species, the labels for 280 species are fixed with posterior probability one, i.e., the same labels are selected for each 280 species for every posterior sample.
The number of distinct labels, i.e., associated with at least one species, is 22 with posterior probability one.

We also calculated the inefficiency factor (IF) which is the ratio of the numerical variance of the estimate from the MCMC samples relative to that from hypothetically uncorrelated samples. It is defined as $1+2\sum_{s=1}^{\infty}\rho_{s}$ where $\rho_{s}$ is the sample autocorrelation at lag $s$.
It suggests the relative number of correlated draws necessary to attain the same variance of the posterior mean from the uncorrelated draws \citep{Chib(01)}. The IFs for parameters are $53 \sim 140$.
Since we retain 20,000 samples as posterior draws, we preserve at least $20,000/140\approx 142$ samples from the stationary distribution. The computational time for 40,000 iterations with 5 factors is 3,211 minutes.

\begin{figure}[ht]
  \begin{center}
   \includegraphics[width=8cm]{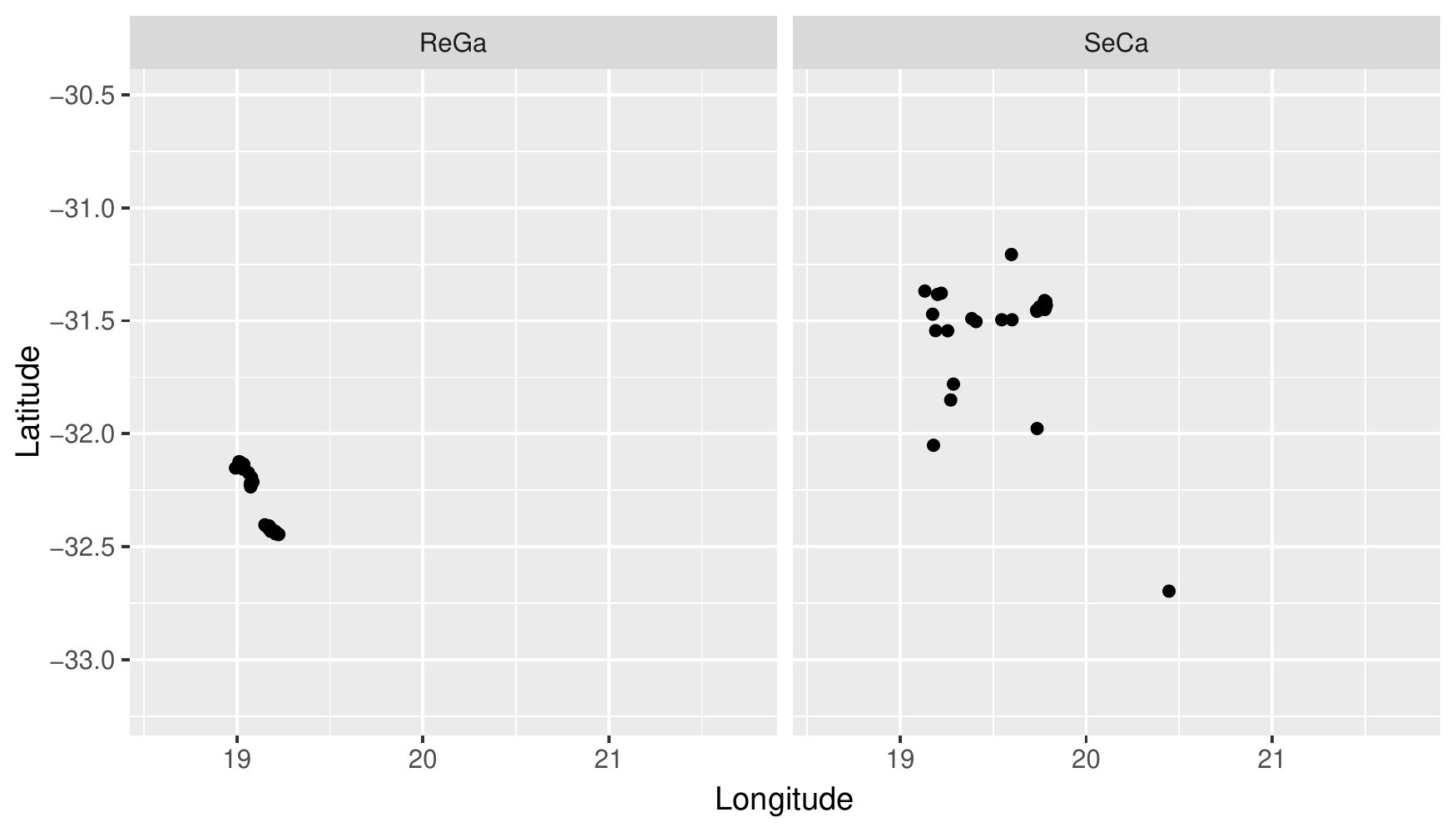}
  \end{center}
  \caption{The distribution of ReGa (left) and Seca (right). }
  \label{fig:Real1-Loc}
\end{figure}

We pick up two species, as discussed in \ref{sec:I}, which share the same label, in particular a label arising from a large negative, hence influential, $\mathbf{W}\mathbf{\Lambda}^{T}$.
One is {\it Restio gaudichaudianus} (ReGa) which shows large absolute values of $\mathbf{X}\mathbf{B}_{l}^{T}$ and the other is {\it Senecio cardaminifolius} (SeCa) which shows small absolute values.
Figure \ref{fig:Real1-Loc} shows the distribution of Rega and Seca. Both species show very different distribution patterns. Rega concentrates in a small southwest area.

\begin{figure}[ht]
  \begin{center}
   \includegraphics[width=12cm]{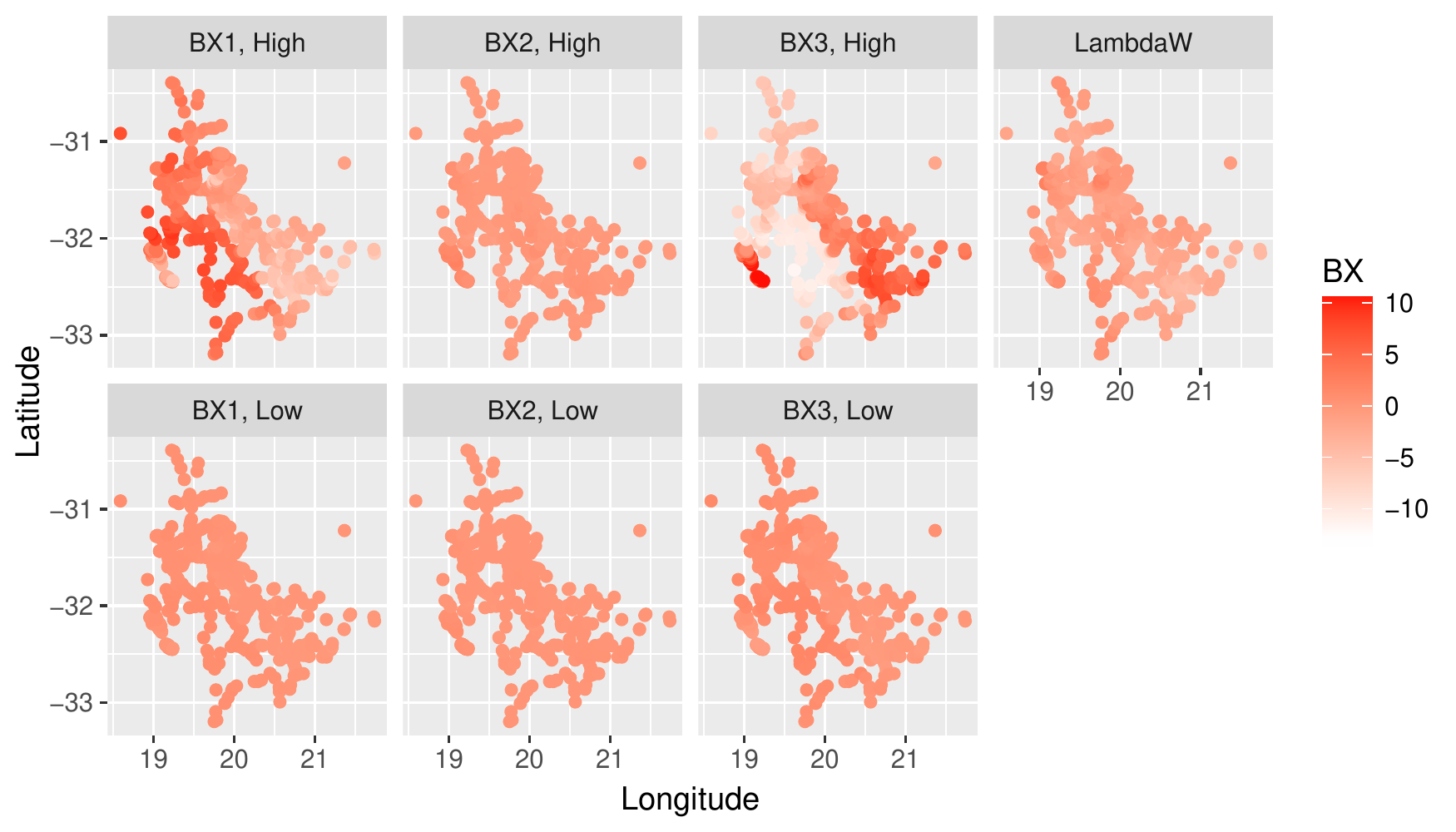}
  \end{center}
  \caption{Estimated $\mathbf{X}\mathbf{B}_{l}^{T}$ and $\mathbf{W}\mathbf{\Lambda}_{l}^{T}$ for ReGa (high, top) and Seca (low, bottom). }
  \label{fig:Real1-Beta}
\end{figure}

\begin{figure}[ht]
  \begin{center}
   \includegraphics[width=12cm]{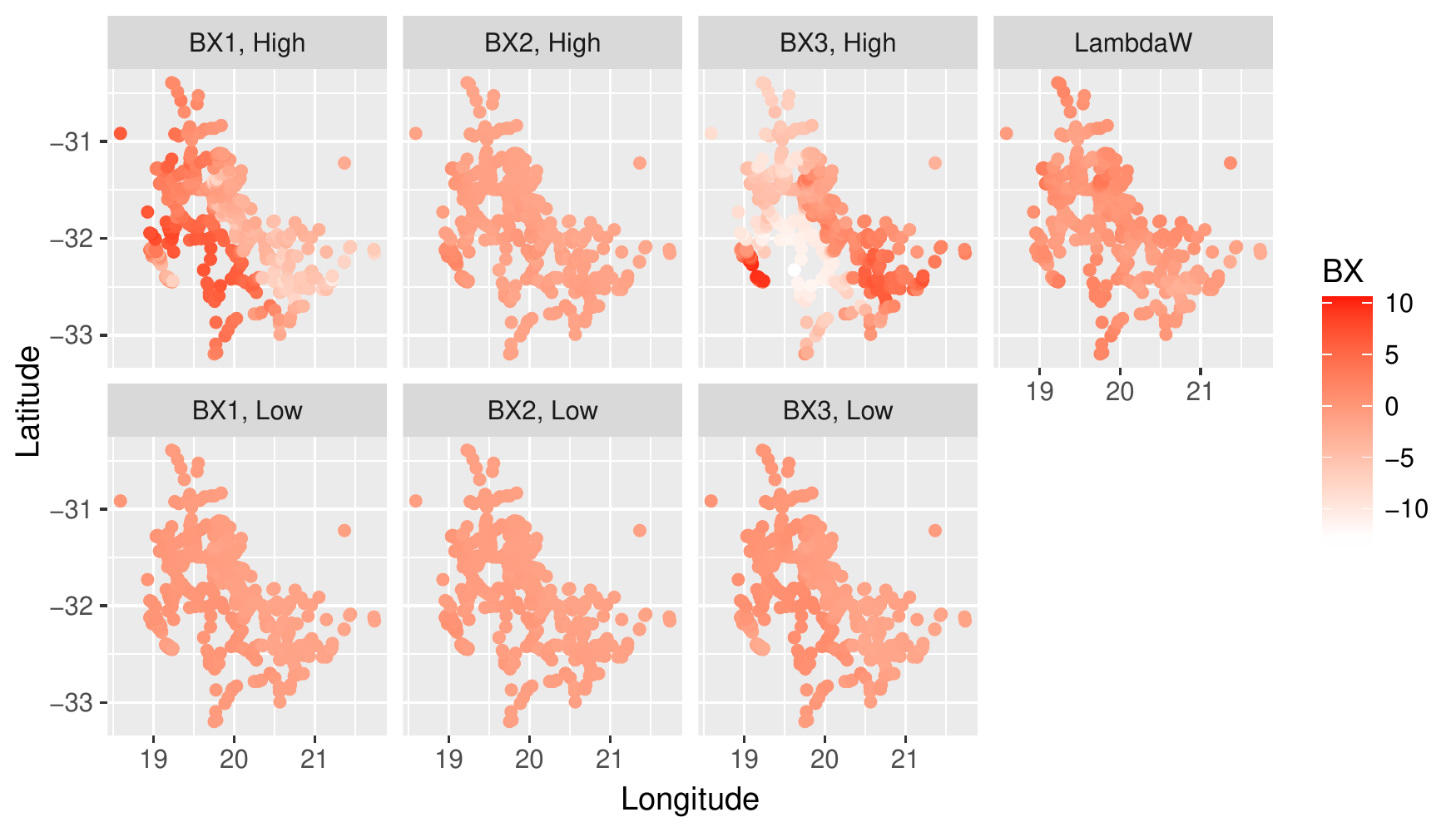}
  \end{center}
  \caption{Estimated orthogonalized $\mathbf{X}\mathbf{B}_{l}^{*T}$ and $\mathbf{W}^{*}\mathbf{\Lambda}_{l}^{T}$ for ReGa (high, top) and Seca (low, bottom). }
  \label{fig:Real1-BetaOrth}
\end{figure}

Figure \ref{fig:Real1-Beta} shows the estimation result of $\mathbf{X}\mathbf{B}_{l}^{T}$ and $\mathbf{W}\mathbf{\Lambda}_{l}^{T}$.
Since they share the same label, $\mathbf{W}\mathbf{\Lambda}_{l}^{T}$ is the same for both species.
For ReGa, $\mathbf{X}\mathbf{B}_{l}^{T}$ reveals larger variation than that for Seca. $\mathbf{W}\mathbf{\Lambda}_{l}^{T}$ shows relatively negative values which exert much influence on the presence probability of Seca.
Figure \ref{fig:Real1-BetaOrth} demonstrates the estimation results for the orthogonalized versions $\mathbf{X}\mathbf{B}_{l}^{*T}$ and $\mathbf{W}^{*}\mathbf{\Lambda}_{l}^{T}$ as defined in Section \ref{sec:I}.
Although the difference is small, the surface of $\mathbf{W}^{*}\mathbf{\Lambda}_{l}^{T}$ has larger positive values than  $\mathbf{W}\mathbf{\Lambda}_{l}^{T}$.
However, the figure suggests that spatial confounding effects are relatively small.

\begin{figure}[ht]
  \begin{center}
   \includegraphics[width=8cm]{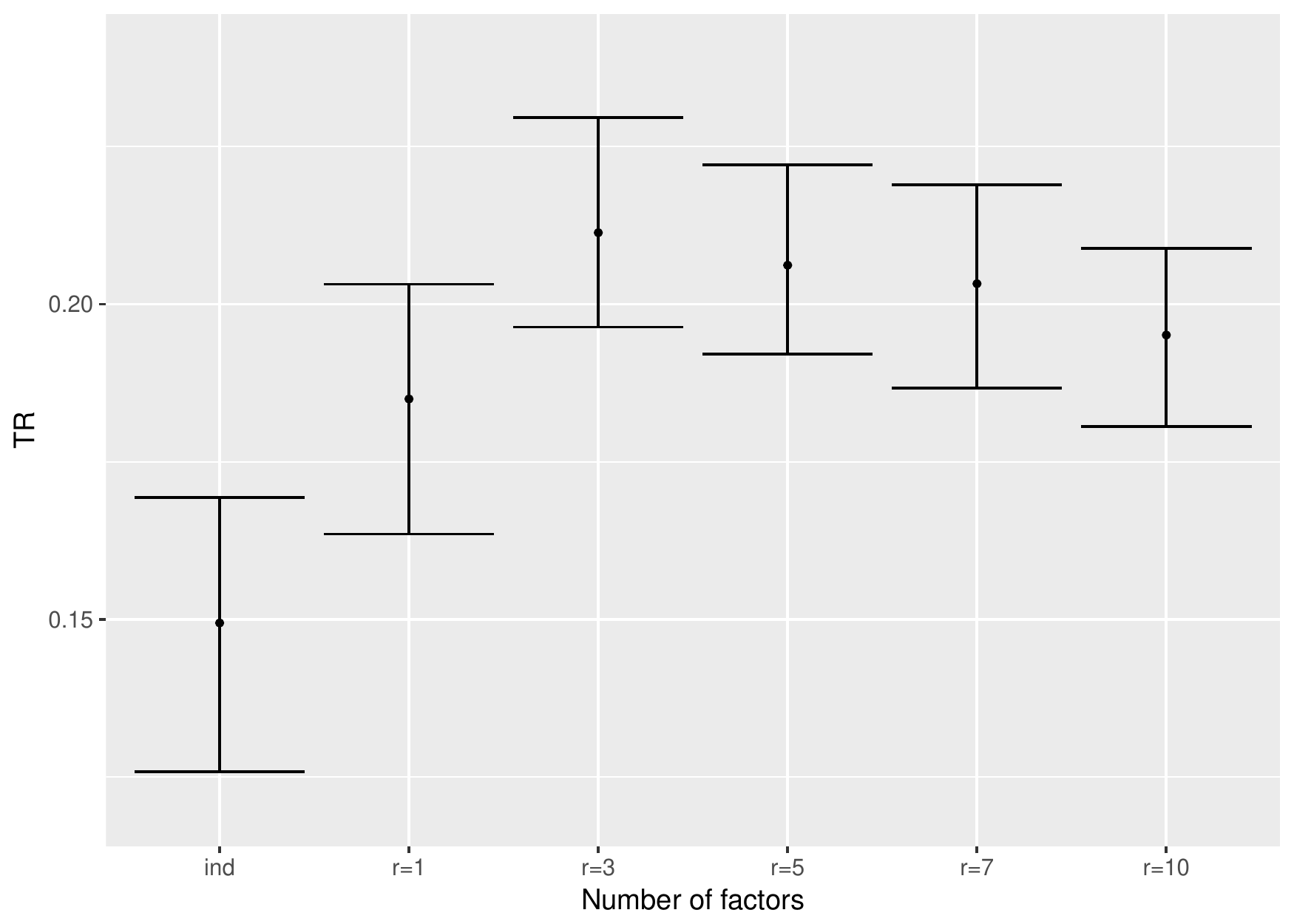}
  \end{center}
  \caption{TR measure for each number of factors. }
  \label{fig:Real1-Sensitivity}
\end{figure}

Next, we investigate the predictive performance of our model.
For a sensitivity check with respect to the number of factors, Figure \ref{fig:Real1-Sensitivity} shows the TR measure for the independence model with 5 factors (first boxplot) and spatial models with different number of factors.
The figure suggests the spatial model with $r=3$ factors shows best performance while the spatial model with 5 factors is similar. Both models show better predictive performance than the independence model with 5 factors.
Also, the models with more factors do not improve performance.

Lastly, we compare the predictive performance between our models and the stacked ``independence'' model. Here, the independence model means that spatial random effects are introduced independently across species. Hence, the stacked independence model incorporates spatial dependence but not dependence among species.
We calculate the conditional TR measure, denoted by $TR_{k|Y^{(l)}=1}$ and $TR_{k|Y^{(l)}=0}$ if we condition on species $l$ being present or absent, respectively, as investigated in \cite{Taylor-Rodriguezetal(17)}.
We illustrate this conditional TR measure at $134$ held out locations by conditioning on the presence-absence state of {\it Aridaria noctiflora} (ArNo) and obtain the posterior probability of presence for {\it Pteronia glomerata} (PtGl). These species share the same label with posterior probability one, and the posterior mean correlation between the two species is 0.4011, which is relatively high.
We calculate $TR_{PtGl|Y_{ArNo}=1}$ and $TR_{PtGl|Y_{ArNo}=0}$ under both the joint model with $r=5$ and the stacked independence model (Table \ref{tab:TRpair}).  The joint model shows better validation performance.

\begin{table}[ht]
\caption{Tjur R for PtGl conditional on ArNo at 134 held out locations}
\label{tab:TRpair}
\centering
\begin{tabular}{lccccccc}
\hline
\hline
 &  & \multicolumn{2}{c}{PtGl}  & & \multicolumn{2}{c}{$TR_{PtGl|ArNo}$}  \\
  &     & 0 & 1  & & Independent & Joint   \\
\hline
\multirow{2}{*}{ArNo} & 0 & $n_{00}=100$ & $n_{01}=12$ &  & 0.2263 & 0.2523  \\
& 1 & $n_{10}=17$ & $n_{11}=5$ &  & 0.2574 & 0.2874   \\
\hline
\hline
\end{tabular}
\end{table}


\section{Summary and future work}\label{sec:SFW}
We have proposed spatial joint species distribution modeling with Dirichlet process dimension reduction for the factor loading matrix.  The former enables dependence across spatial locations, the latter enables the dependence across species.
We show that introduction of spatial dependence into the factors improves out-of-sample predictive performance over the study region under both continuous and binary species response with both simulated and real data.

Future work will consider extending our model to handle more challenging responses.  For instance, we often observe a compositional data response vector, a response which lies on a simplex in $R^{S}$ dimensional space but allows for point masses at $0$'s.  Another challenge is the case of a large number of spatial locations, for instance, at continental scales resulting in perhaps $n\approx 10^{6}$.
In this case, we will explore recently developed sparse Gaussian processes approximation, e.g., the nearest neighbor Gaussian processes \citep[NNGP,][]{Dattaetal(16a)} or the multiresolution Gaussian processes \citep[MGP,][]{Katzfuss(17)}).
Another direction is a more detailed investigation of the effects of additional decay parameters with regard to the covariance matrices of the spatial factors. \cite{RenBanerjee(13)} allow different decay parameters for spatial factor models, $\phi_{h}$ for $h=1,\ldots, r$ with Gaussian predictive process approximation by \cite{Banerjeeetal(08)}. Without some approximation of the Gaussian processes, inference with different decay parameters requires us to compute matrix factorizations $r$ times for sampling $\phi_{h}$ for $h=1,\ldots, r$
which is computationally demanding even when the number of locations is moderate. Again, the NNGP or MGP approach may be useful for this situation.

\par
\section*{Acknowledgements}
The computational results are obtained by using Ox version 7.1 \citep{Doornik(07)}. The work of the first and third authors was supported, in part, by federal grants NSF/DMS 1513654, NSF/IIS 1562303 and NIH/NIEHS 1R01ES027027. The authors
thank Matthew Aiello-Lammens and John A. Silander, Jr. for providing the Cape Floristic
Region data as well as for motivation and useful conversations regarding the problem.
\par

\appendix
\section*{Appendix}
\section{Details of model fitting}\label{app:A1}
\noindent{\bf Sampling $\mathbf{B}$} \\
Let $\bm{x}_{i}$ be a $p\times 1$ location dependent covariate vector, which is assumed common for the $l=1,\ldots, S$ species.
For $\mathbf{B}_{l}$, we have $\mathbf{B}_{l}\sim \mathcal{N}(\bm{\mu}_{\mathbf{B}_{l}}, \mathbf{\Sigma}_{\mathbf{B}})$
where
\begin{align}
\bm{\mu}_{\mathbf{B}_{l}}=\mathbf{\Sigma}_{\mathbf{B}}\frac{1}{\sigma_{\epsilon}^2}\mathbf{X}^{T}(\mathbf{U}^{(l)}-\mathbf{W}(\mathbf{Z}^{T}\mathbf{Q}(\bm{k})^{T})^{(l)}), \quad \mathbf{\Sigma}_{\mathbf{B}}=\biggl(\frac{\mathbf{X}^{T}\mathbf{X}}{\sigma_{\epsilon}^2}+\frac{1}{c}\mathbf{I}_{S}\biggl)^{-1}
\end{align}
with $\mathbf{U}^{(l)}$ is the $l$-th column of matrix $\mathbf{U}$ and $(\mathbf{Z}^{T}\mathbf{Q}(\bm{k})^{T})^{(l)}$  the $l$-th column of matrix $\mathbf{Z}^{T}\mathbf{Q}(\bm{k})^{T}$.
\\
\noindent{\bf Sampling $\mathbf{Z}$} \\
Sampling $\mathbf{Z}$ employs almost the same algorithm as in \cite{Taylor-Rodriguezetal(17)}.
In our case, the first row of $\mathbf{\Lambda}$ is positive, we set $\bm{Z}_{1}$ as the first row of $\mathbf{\Lambda}$.
For $j=1$,
\begin{itemize}
\item let $S_{1}=\{l=1,\ldots, S, \text{s.t.} k_{l}=1\}$ and let $|S_{1}|$ denote the cardinality of $S_{1}$. Using these definitions the full conditional distribution for $\bm{Z}_{1}$ is given by
$\bm{Z}_{1}\sim \mathcal{TN}_{r}(\bm{\mu}_{\bm{Z}_{1}}, \mathbf{\Sigma}_{\bm{Z}_{1}})$
where $\mathcal{TN}_{r}$ is multivariate truncated normal distribution defined on $(0, \infty)^{r}$ and
\begin{align}
\bm{\mu}_{\bm{Z}_{1}}=\mathbf{\Sigma}_{\bm{Z}_{1}}\mathbf{W}^{T} \frac{1}{\sigma_{\epsilon}^2}\sum_{l\in S_{1}} (\mathbf{U}^{(l)}-\mathbf{X}\mathbf{B}_{l}^{T}), \quad
\mathbf{\Sigma}_{\bm{Z}_{1}}=\biggl(\frac{|S_{1}|}{\sigma_{\epsilon}^2}\mathbf{W}^{T}\mathbf{W}+\mathbf{D}_{\bm{Z}}^{-1}\biggl)^{-1}
\end{align}
\end{itemize}
The full conditional for other rows of $\mathbf{Z}$ depends on whether or not the row considered was chosen to be at least one row from $\mathbf{\Lambda}$,
For $j=2,\ldots, N$
\begin{itemize}
\item[1.] If $j\notin \bm{k}$, sample $\bm{Z}_{j}\sim \mathcal{N}_{r}(\bm{0}, \mathbf{D}_{\bm{Z}})$.
\item[2.] Otherwise, let $S_{j}=\{l=1,\ldots, S, \text{s.t.} k_{l}=j\}$ and let $|S_{j}|$ denote the cardinality of $S_{j}$. Using these definitions the full conditional distribution for $\bm{Z}_{j}$ is given by
$\bm{Z}_{j}\sim \mathcal{N}_{r}(\bm{\mu}_{\bm{Z}_{j}}, \mathbf{\Sigma}_{\bm{Z}_{j}})$
where
\begin{align}
\bm{\mu}_{\bm{Z}_{j}}=\mathbf{\Sigma}_{\bm{Z}_{j}}\mathbf{W}^{T} \frac{1}{\sigma_{\epsilon}^2}\sum_{l\in S_{j}} (\mathbf{U}^{(l)}-\mathbf{X}\mathbf{B}_{l}^{T}), \quad
\mathbf{\Sigma}_{\bm{Z}_{j}}=\biggl(\frac{|S_{j}|}{\sigma_{\epsilon}^2}\mathbf{W}^{T}\mathbf{W}+\mathbf{D}_{\bm{Z}}^{-1}\biggl)^{-1}
\end{align}
\end{itemize}
with $\mathbf{B}_{l}$ the $l$-th row of matrix $\mathbf{B}$.
\\
\noindent{\bf Sampling $\mathbf{W}$} \\
Sampling $\mathbf{W}$ requires the matrix factorization for $n$-dimensional covariance matrices.
For $h=1,\ldots, r$,
\begin{align}
[\mathbf{W}^{(h)}|\cdot]\propto \prod_{i=1}^{n}\exp\biggl(-\frac{1}{2\sigma_{\epsilon}^2}\|\bm{U}_{i}-\mathbf{B}\bm{x}_{i}-\mathbf{Q}(\bm{k})\mathbf{Z}\bm{w}_{i}\|^2 \biggl)\times \exp\biggl(-\frac{1}{2}\mathbf{W}^{(h)T}\mathbf{C}_{\phi}^{-1} \mathbf{W}^{(h)}\biggl) \nonumber \\
\end{align}
Although Gibbs sampling is available, $\mathcal{O}(n^3)$ computational time is required.

Let $\mathbf{Z}^{(h)}$ be $h$-th column vector of $\mathbf{Z}$, $\mathbf{Z}^{(-h)}$ and $\mathbf{W}^{(-h)}$ be remaining matrices after deleting $\mathbf{Z}^{(h)}$ and $\mathbf{W}^{(h)}$, respectively.
The full conditional is
\begin{align}
[\mathbf{W}^{(h)}|\cdot]&\propto \exp\biggl(-\frac{1}{2\sigma_{\epsilon}^2}\biggl(\mathbf{U}-\mathbf{X}\mathbf{B}^{T}-\mathbf{W}\mathbf{Z}^{T}\mathbf{Q}(\bm{k})^{T} \biggl)^{T} \biggl(\mathbf{U}-\mathbf{X}\mathbf{B}^{T}-\mathbf{W}\mathbf{Z}^{T}\mathbf{Q}(\bm{k})^{T}\biggl) \biggl) \nonumber \\
&\times \exp\biggl(-\frac{1}{2}\mathbf{W}^{(h)T}\mathbf{C}_{\phi}^{-1} \mathbf{W}^{(h)}\biggl) \nonumber \\
&\propto \exp \biggl( -\frac{1}{2\sigma_{\epsilon}^2} \biggl( \mathbf{U}-\mathbf{X}\mathbf{B}^{T}-\mathbf{W}^{(-h)}\mathbf{Z}^{(-h)T}\mathbf{Q}(\bm{k})^{T}-\mathbf{W}^{(h)}\mathbf{Z}^{(h)T}\mathbf{Q}(\bm{k})^{T} \biggl)^{T} \nonumber \\
&\times \biggl(\mathbf{U}-\mathbf{X}\mathbf{B}^{T}-\mathbf{W}^{(-h)}\mathbf{Z}^{(-h)T}\mathbf{Q}(\bm{k})^{T}-\mathbf{W}^{(h)}\mathbf{Z}^{(h)T}\mathbf{Q}(\bm{k})^{T} \biggl) \biggl) \times \exp\biggl(-\frac{1}{2}\mathbf{W}^{(h)T}\mathbf{C}_{\phi}^{-1} \mathbf{W}^{(h)}\biggl) \nonumber \\
&=\mathcal{N}(\bm{\mu}_{w_{h}}, \mathbf{\Sigma}_{w_{h}})
\end{align}
where
\begin{align}
\bm{\mu}_{w_{h}}&=\mathbf{\Sigma}_{w_{h}}\frac{1}{\sigma_{\epsilon}^2}\biggl(\mathbf{U}-\mathbf{X}\mathbf{B}^{T}-\mathbf{W}^{(-h)}\mathbf{Z}^{(-h)T}\mathbf{Q}(\bm{k})^{T} \biggl)\mathbf{Q}(\bm{k})\mathbf{Z}^{(h)} \\
\mathbf{\Sigma}_{w_{h}}&=\biggl(\mathbf{C}_{\phi}^{-1}+\frac{\|\mathbf{Z}^{(h)T}\mathbf{Q}(\bm{k})^{T}\|^2}{\sigma_{\epsilon}^2}\mathbf{I}_{n} \biggl)^{-1}
\end{align}
\\
\noindent{\bf Sampling $\phi$} \\
The full conditional distribution for $\phi$ is
\begin{align}
|\mathbf{C}_{\phi}|^{-\frac{1}{2}}\exp\biggl(-\frac{1}{2}\mathbf{W}^{(h)T}\mathbf{C}_{\phi}^{-1} \mathbf{W}^{(h)}\biggl)\mathbf{I}(\phi_{min}<\phi<\phi_{max})
\end{align}
We implement a Metropolis-Hastings algorithm.
\\
\noindent{\bf Sampling $\bm{k}$} \\
For the vector of labels $\bm{k}$, the full conditional distribution is
$[\bm{k}|\cdot]=\prod_{l=1}^{S}\biggl(\sum_{j=1}^{N}p_{l,j} \delta_{j}(k_{l})\biggl)$
with
\begin{align}
p_{l,j} \propto p_{j}\times \exp\biggl(-\frac{1}{2\sigma_{\epsilon}^2}\|\mathbf{U}^{(l)}-\mathbf{X}\mathbf{B}_{l}^{T}-\mathbf{W}\bm{Z}_{j}\|^2 \biggl)
\end{align}
\\
\noindent{\bf Sampling $\bm{p}$} \\
The full conditional distribution for $\bm{p}$, given conjugacy of the $\mathcal{GD}$ distribution with multinomial sampling, the draws of $\bm{p}$ are
\begin{align}
p_{1}&=\xi_{1},  \\
p_{j}&=(1-\xi_{1})\cdots(1-\xi_{j-1})\xi_{j}, \quad \text{for} \quad j=2,3,\ldots, N-1 \\
p_{N}&=1-\sum_{j=1}^{N-1} p_{j},
\end{align}
with $\xi_{j}\sim \text{Beta}(\frac{\alpha}{N}+\sum_{l=1}^{S}I_{(k_{l}=j)}, \frac{N-1}{N}\alpha+\sum_{s=j+1}^{N}\sum_{l=1}^{S}I_{(k_{l}=s)})$ for $j=1,\ldots, N-1$.
\\
\noindent{\bf Sampling $\sigma_{\epsilon}^2$} \\
By conjugacy of the prior for $\sigma_{\epsilon}^2$ with the normal likelihood, the full conditional distribution is
\begin{align}
\sigma_{\epsilon}^2 \sim \mathcal{IG} \biggl(\frac{nS+a}{2}, \frac{\sum_{i=1}^{n}\|\bm{U}_{i}-\mathbf{B}\bm{x}_{i}-\mathbf{Q}(\bm{k})\mathbf{Z}\bm{w}_{i} \|^2+b }{2} \biggl)
\end{align}
\\
\noindent{\bf Sampling $\mathbf{D}_{\bm{Z}}$} \\
\begin{align}
\mathbf{D}_{\bm{Z}}\sim \mathcal{IW}\biggl(\mathbf{D}_{\bm{Z}}|2+r+N-1, \mathbf{Z}^{T}\mathbf{Z}+4\text{diag}\biggl(\frac{1}{\eta_{1}}, \ldots, \frac{1}{\eta_{r}} \biggl) \biggl)
\end{align}

\markboth{\hfill{\footnotesize\rm Shinichiro Shirota, Sudipto Banerjee and Alan E. Gelfand} \hfill}
{\hfill {\footnotesize\rm Spatial Species Distribution Modeling} \hfill}

\bibhang=1.7pc
\bibsep=2pt
\fontsize{9}{14pt plus.8pt minus .6pt}\selectfont
\renewcommand\bibname{\large \bf References}
\expandafter\ifx\csname
natexlab\endcsname\relax\def\natexlab#1{#1}\fi
\expandafter\ifx\csname url\endcsname\relax
  \def\url#1{\texttt{#1}}\fi
\expandafter\ifx\csname urlprefix\endcsname\relax\def\urlprefix{URL}\fi

\bibliographystyle{chicago}
\bibliography{SP}
\vskip .65cm
\noindent
Department of Biostatistics, University of California, Los Angeles. 650 Charles E. Young Drive South Los Angeles, CA 90095-1772
\vskip 2pt
\noindent
E-mail: shinichiro.shirota@gmail.com
\vskip 2pt

\noindent
Department of Statistics, Duke University, Durham, NC 27708-0251
\vskip 2pt
\noindent
E-mail: alan@duke.edu
\vskip 2pt

\noindent
Department of Biostatistics, University of California, Los Angeles. 650 Charles E. Young Drive South Los Angeles, CA 90095-1772
\vskip 2pt
\noindent
E-mail: sudipto@ucla.edu

\end{document}